\documentstyle[11pt,epsfig]{article}

\renewcommand{\theequation}{\arabic{section}.\arabic{equation}}

\def\laq{~\raise 0.4ex\hbox{$<$}\kern -0.8em\lower 0.62
ex\hbox{$\sim$}~}
\def\gaq{~\raise 0.4ex\hbox{$>$}\kern -0.7em\lower 0.62
ex\hbox{$\sim$}~}
\def\vp{\varphi}
\def \vpb {{\overline {\vp}}}

\def \pa {\partial}

\begin{document}

\begin{titlepage}

\begin{flushright}
CERN-PH-TH/2004-102\\
\end{flushright}

\vspace{0.8cm}

\begin{center}

\huge{Heating up the cold bounce}

\vspace{0.8cm}

\large{Massimo Giovannini}

\normalsize
\vspace{.1in}

{\sl Department of Physics, Theory Division, CERN 1211 Geneva 23, Switzerland }

\vspace*{1cm}

\begin{abstract}
Self-dual string cosmological models provide an effective example of bouncing solutions
where a phase of accelerated contraction smoothly evolves into  an epoch of  decelerated 
Friedmann--Robertson--Walker expansion dominated by the dilaton. While the transition 
to the expanding regime occurs at sub-Planckian curvature scales, 
the Universe emerging after the bounce is cold, with sharply growing gauge coupling.
However, since massless gauge bosons (as well as other massless fields) are super-adiabatically amplified,  
the energy density of the maximally amplified modes re-entering the horizon 
after the bounce  can efficiently heat the Universe. As a consequence 
the gauge coupling reaches a constant value, which can still be perturbative.
\end{abstract}
\end{center}
\end{titlepage}

\newpage

\parskip 0.2cm

\renewcommand{\theequation}{1.\arabic{equation}}
\setcounter{equation}{0}
\section{Formulation of the problem} 
One of the inspiring symmetries of the pre-big bang scenario is  the   
scale factor duality (SFD),  allowing the connection of 
different cosmological solutions of the low-energy string 
effective action \cite{sfd}. For instance, in the string frame description, accelerated solutions 
with growing curvature are connected via SFD to decelerated 
Friedmann--Robertson--Walker (FRW) models with 
decreasing curvature and decreasing dilaton energy density.
The same solutions can also be described in the so-called 
Einstein frame, where the dilaton is not directly coupled to the 
Einstein--Hilbert term \cite{gv1}. In this case a solution 
exhibiting accelerated contraction evolves into a decelerated FRW expansion.

The connection among different solutions provided by the combined action of SFD 
and time-reversal does not imply an analytical connection of the 
duality-related solutions, as clearly 
noted in the early days of the pre-big bang cosmology \cite{gv2}.
In spite of this, it is known that solutions connecting smoothly two 
duality-related regimes exist in the presence of dilaton potentials which are 
explicitly invariant under the SFD symmetry \cite{gv2,meis,gmv}. Recently its was 
shown \cite{ggv1,ggv2} that these models can be derived from a 
generally covariant action supplemented by a  non-local dilaton potential. In the 
absence of fluid sources and in the Einstein frame metric, these solutions 
smoothly interpolate between a phase of accelerated contraction  
valid for negative values of the conformal-time coordinate and a phase of decelerated 
expansion valid for positive values of the conformal time coordinate. The intermediate 
high-curvature phase is regular and can be interpreted as cold bounce (CB) since 
the Universe is never dominated by radiation, be it before or after the 
bounce. Furthermore, the nature of CB solutions also implies that the 
dilaton field never becomes constant for $\eta \to +\infty $ but 
 is always growing logarithmically. 

A possible way out is to consider the effects 
of the high-frequency (small-scale) modes amplified during the epoch 
preceding the bounce. In the low-energy effective action also 
Abelian gauge bosons may be present and since their coupling to the dilaton field
breaks conformal invariance \cite{ggv3} the various modes 
of the ``photon'' field can be super-adiabatically amplified. 
While inside the horizon the quantum mechanical fluctuations related to these 
modes are adiabatically damped. On the contrary, as the modes get outside the horizon, their 
energy density  is super-adiabatically amplified, thanks to the coupling of the kinetic term of the 
gauge bosons to the dilaton field. When the fluctuations re-enter the horizon, after the bounce, their 
energy density red-shifts as radiation while the energy density of the background, still dominated by 
the dilaton field, red-shifts as stiff matter. Since the energy density of the background 
decreases faster than radiation, there will be a moment where the energy density 
of the radiation background will become 
dominant, providing a mechanism for the gravitational heating of the cold bounce
\cite{paschos}. Well after the bounce the dilaton potential is negligible with 
respect to the dilaton kinetic term; this means that when radiation 
starts dominating the dilaton field will be driven to a constant asymptotic value for $\eta \to +\infty$.

The idea that back-reaction effects can produce dynamically a radiation-dominated background  
has been previously discussed within different approaches, but not specifically in connection 
with the problem of dilaton stabilization. 
The idea of a possible gravitational reheating of the Universe was  pointed out by 
Parker \cite{parker1,parker2} and also, in a related perspective, by Grishchuk \cite{gr1}.
Lukash and Starobinsky \cite{LS}  considered similar effects but paid particular 
attention to the process of isotropization of anisotropic solutions by particle production.
In more recent times, Ford \cite{ford1} discussed the problem of production of slightly 
non-conformally coupled fields in the case of inflationary Universes. 
Even more recently Peebles and Vilenkin \cite{pv1} used the mechanism of gravitational reheating in  
the case of quintessential  inflation (see also \cite{mgqi}). 

In the following the different steps outlined in the previous paragraphs looked into.
In Section 2 the considerations leading to cold bounces  are  discussed
 both in the String and in the Einstein frame. In Section 3 the 
amplification of Abelian gauge bosons are accurately computed. Section 4 
describes the gravitational reheating of the cold bounces. The concluding remarks  are 
presented in Section 5. The Appendices contain useful technical results 
supporting and complementing the results of Sections 2 and 4.

\renewcommand{\theequation}{2.\arabic{equation}}
\setcounter{equation}{0}
\section{Cold bounces in the string and Einstein frames} 
The string frame action leading to cold-bounce solutions  \cite{ggv1,ggv2}  
can be written as 
\begin{equation}
S = - \frac{1}{2\lambda_{s}^{2}} \int d^{4} x \sqrt{-G} e^{-\varphi}[ 
R + G^{\alpha\beta} \partial_{\alpha} \varphi \partial_{\beta} \varphi 
+ V(\overline{\varphi})] + S_{\rm m}
\label{action1}
\end{equation}
where $\lambda_{s}$ is the string length scale and the metric signature 
is mostly minus \footnote{Greek indices run over the four space-time dimensions, while 
Latin indices take values over the  three-dimensional spatial geometry.}. 
In Eq. (\ref{action1}) $S_{\rm m}$ account for  the possible 
contribution of matter sources.  

The potential $V(\vpb(x))$, a local function of $\vpb$, is instead a
non-local function (yet a scalar under general coordinate transformations)
of the dilaton owing to the definitions:
\begin{equation}
V= V(e^{-\vpb}),  ~~~~~~
e^{-\vpb(x)} =    \int {d^{4}y \over  \lambda_s^2}\sqrt{-G(y)}~ e^{-\vp(y)}
\sqrt{G^{\mu\nu} \pa_\mu \vp (y) \pa_{\nu} \vp(y)}
~\delta\left(\vp (x) -   \vp (y)\right) .
\label{POTdef}
\end{equation}
From the metric $G_{\mu\nu}$, the induced metric
\begin{equation}
\gamma_{\mu\nu} = G_{\mu\nu} - \frac{\partial_{\mu}\vp \partial_{\nu}\vp}{(\partial \vp)^2},  
\label{indmetr}
\end{equation}
and the induced Laplacian
\begin{equation}
\hat{\nabla}^2 \vp = \gamma_{\mu\nu} \nabla^\mu\nabla^\nu \vp,
\label{indlap}
\end{equation}
can be easily defined. Thus, in terms of Eqs. (\ref{indmetr}) and (\ref{indlap}), 
the variation of the action (\ref{action1}) with respect to $G_{\mu\nu}$ and
$\vp$ leads, respectively, to the following two equations:
\begin{equation}
R_{\mu}^{\nu} - \frac{1}{2} \delta_{\mu}^{\nu} + \nabla_{\mu} \nabla^{\nu} \vp + \frac{1}{2} \delta_{\mu}^{\nu}  
\left[ G^{\alpha\beta} \nabla_{\alpha} \vp \nabla_{\beta} \vp  - 2 G^{\alpha\beta} \nabla_{\alpha}\nabla_{\beta} \vp - V\right] - 
 \frac{1}{2} e^{-\vp} 
\sqrt{(\partial \vp)^2} ~\gamma_{\mu}^{\nu} I_1 = \lambda_{s}^2 e^{\varphi} T_{\mu}^{\nu},
\label{genback1}
\end{equation}
\begin{equation}
R + 2 G^{\alpha\beta} \nabla_{\alpha} \nabla_{\beta}\vp - G^{\alpha\beta} \partial_{\alpha}\vp \partial_{\beta} \vp 
 + V - {\pa V\over \pa{\bar{\vp}}} 
+  e^{-\vp}  \frac{\hat{\nabla}^2 \vp}{\sqrt{(\partial \vp)^2}}  I_1 - 
e^{-\vp}  V'   I_2 = 0,
\label{genback2}
\end{equation}
With standard notation, $T_{\mu}^{\nu}$ is the 
energy--momentum tensor of the sources, while the integrals 
\begin{eqnarray}
&&
I_1 = {1\over \lambda_s^3}\int d^{4}y \sqrt{-G(y)}~V'(e^{-\vpb(y)}) 
~\delta\left(\vp (x) -   \vp (y)\right) , 
\nonumber\\
&&
I_2 = {1\over \lambda_s^3} \int d^{4} y \sqrt{-G(y)}~ 
\sqrt{\pa_\mu \vp (y) \pa^\mu \vp(y)}
~\delta'\left(\vp (x) -   \vp (y)\right) \; ,
\end{eqnarray} 
arise directly from the standard variational procedure.

For a homogeneous, isotropic and spatially flat background, 
\begin{equation} 
g_{00}=1,~~~~~~~g_{ij}=-a_{s}^2(t)\delta_{ij},~~~~~~~\varphi=\varphi(t),~~~~~~~T_{0}^{0} = \rho_{s}(t),~~~~~~~ 
T_{i}^{j} = - p_{s}(t) \delta_{i}^{j},
\label{homback}
\end{equation}
the evolution equations (\ref{genback1}) and (\ref{genback2}) greatly simplify
and become perfectly local in time. In particular, from Eq. (\ref{POTdef})  
\begin{equation}
e^{-\vpb}= e^{-\vp}a_{s}^3,
\end{equation}
having absorbed into $\vp$ the dimensionless constant $-\ln 
(\int d^3y/\lambda_s^3)$, associated with the (finite) comoving spatial volume. 
In the class of backgrounds specified by Eq. (\ref{homback}), 
the time and space components of Eq. (\ref{genback1}), and the dilaton equation (\ref{genback2}), 
lead  to the following set of equations 
\begin{eqnarray}
&& \dot{\overline{\varphi}}^2 - 3 H_s^2 - V = e^{\vpb} \overline{\rho}_{s} ,
\label{b1}\\
&& \dot{H}_s = \dot{\overline{\varphi}} H_s + \frac{1}{2} e^{\vpb}  \overline{p}_{s},
\label{b2}\\
&& 2 \ddot{\overline{\varphi}} - \dot{\overline{\varphi}}^2 - 3 H_s^2 + V - \frac{\partial V}{\partial \overline{\varphi}} =0,
\label{b3}\\
&& \dot{\overline{\rho}_{s}} + 3 H \overline{p}_{s} =0,
\label{b4}
\end{eqnarray}
where the dot denotes derivation with respect to the cosmic time in the 
string frame and where $H_{s} = \dot{a}_{s}/a_{s}$. In Eqs. (\ref{b1})--(\ref{b4}) 
the following quantities have been defined:
\begin{equation}
\overline{\rho}_{s} = a_s^3 \rho_{s}, ~~~~~\overline{p}_{s} = a_s^3 p_{s}.
\end{equation}
In the string frame (and in the absence of fluid sources)  the  system of Eqs. (\ref{b1})--(\ref{b3}) 
admits various cold-bounce solutions and some  examples are provided in the Appendix (see also \cite{ggv2}).
An interesting case is given, for instance, by the  following  solution:
\begin{eqnarray}
&& V (\overline{\vp}) = - V_0 e^{ 4\overline{\vp} },
\label{pot1}\\
&&a_{s}(t) = a_0 \biggl[ \tau + \sqrt{\tau^2 + 1}\biggr]^{1/\sqrt{3}},
\label{scalefactor1}\\
&& \overline{\vp} = \varphi - 3 \ln a_s(t) =- \frac{1}{2} \log{ ( 1 + \tau^2)} + \vp_{0},
\label{dilaton1}
\end{eqnarray}
where
\begin{equation}
\tau = \frac{t}{t_0}, ~~~~~~~~~~~~ t_0 = \frac{e ^{- 2 \vp_0}}{ \sqrt{V_{0}}}.
\end{equation}
The value of $t_0^{-1}$ sets the maximal value of the Hubble barameter, $H_s(t)$, at the bounce.
The value of $\varphi_0$ sets the typical value of the  dilaton at the bounce.

Using Eq. (\ref{scalefactor1}), it can be checked that 
$a_{s}(t) = a_{s}^{-1}(-t)$ as implied by 
the  invariance of the solution under scale factor duality. Other duality-invariant solutions 
of the system of Eqs. (\ref{b1})--(\ref{b4}) are reported in  Appendix  together with a more 
extensive discussion of the cold-bounce solution.

The conformal 
time coordinate is the same in both Einstein and string frames \cite{gv1}. Thus, in order to swiftly perform the correct 
transition from string to Einstein frame, it is appropriate to rewrite the system of Eqs. (\ref{b1})--(\ref{b4}) 
in the conformal time coordinate, i.e.
\begin{equation}
a_{s}(\eta_{s}) d\eta_s = d t_{s}.
\end{equation} 
After this coordinate change, Eqs. (\ref{b1})--(\ref{b4}) can be written as:
\begin{eqnarray}
&& {\vp'}^2 + 6 {\cal H}_{s}^2 - 6 {\cal H}_{s} \vp' = V a_{s}^2 + a^2 e^{\vp} \rho_{s} ,
\label{b1in}\\
&& {\cal H}_{s}' = {\cal H}_{s} \varphi' - 2 {\cal H}_{s}^2 + \frac{e^{\varphi /2}}{2} a_{s}^2 p_{s},
\label{b2in}\\
&& 2 \vp'' + 4 {\cal H}_{s} \vp' - 6 {\cal H}_{s}' - 6 {\cal H}_{s}^2 + V - \frac{\partial V}{\partial\vp} =0,
\label{b3in}\\
&& \rho_{s}' + 3 {\cal H}_{s} ( \rho_{s} + p_{s}) =0,
\label{b4in}
\end{eqnarray}
where ${\cal H}_{s} = a_{s}'/a_{s}$ and the prime denotes derivation with respect to $\eta_{s}$.
As already mentioned, the dilaton and the conformal time coordinate do not change in the 
transition from Einstein to string frames 
\begin{equation}
\eta_{e} = \eta_{s}=\eta,~~~~~\vp_{e} = \vp_{s}=\vp,
\end{equation}
where $\eta$ and $\vp$ denote the common values of the conformal time coordinate and of the dilaton in both frames.
The scale factor, the Hubble parameter, and the energy and pressure densities 
do change in the transition from string to Einstein frame:
\begin{equation}
a_{s} = e^{\varphi/2} a,~~~~~~ {\cal H}_{s} = {\cal H} + \frac{\varphi'}{2}, ~~~~~~~\rho_{s}= e^{-2\varphi} \rho, ~~~~~~~
p_{s} = e^{-2\varphi} p,
\label{transition}
\end{equation}
where $a$, ${\cal H}$, $\rho$ and $p$ are the Einstein frame quantities while $\varphi$ and $\eta$ will 
be, respectively, the common values of the dilaton and of  the conformal time coordinate in both frames.

Applying the transformations given in (\ref{transition}), Eqs. (\ref{b1in}) and (\ref{b4in}) lead to the Hamiltonian constraint
and to the conservation equation:
\begin{eqnarray}
&& 6  {\cal H}^2 = \rho a^2 +{1\over 2} { \vp'}^2  + e^{\vp}V a^2.
\label{be1}\\
&& \rho' + 3 {\cal H} (\rho + p) - \frac{\varphi'}{2} ( \rho - 3 p) =0.
\label{be4}
\end{eqnarray}
Equations (\ref{b2in}) and (\ref{b3in}) lead, respectively, to
\begin{eqnarray}
&& \varphi'' + 2 {\cal H} \varphi' + 2 {\cal H}' + 4 {\cal H}^2 = pa^2,
\label{bin5}\\
&& \varphi'' +  2 {\cal H} \varphi' + \frac{\partial V}{\partial\varphi} - V + 6 {\cal H}' + 6 {\cal H}^2 - \frac{{\varphi'}^2}{2} =0.
\label{bin6}
\end{eqnarray}
By linear combination of Eqs. (\ref{bin5}) and (\ref{bin6}), the two remaining equations of the system can be obtained, namely:
\begin{eqnarray}
&& 4 {\cal H}'+2 {\cal H}^2 = -p a^2 - \left( \frac{{\vp'}^2}{2}-e^{\vp}
V a^2\right)- e^{\vp}{\partial V\over \partial \vpb}a^2, 
\label{be2}\\
&&
 \varphi^{\prime \prime} +2 {\cal H}  \vp'  
+{1\over 2}(\rho -3 p)a^2
+e^{\vp}
\left(Va^2-{1\over  2}{\partial V\over \partial \vpb}a^2\right)
=0.
\label{be3}
\end{eqnarray}
Equations (\ref{be1})--(\ref{be4}), and (\ref{be2})--(\ref{be3}) 
form a closed set of equations whose solutions will now be analysed first in the absence 
of fluid sources. 

The Einstein frame equations can also be obtained, in generic $d$ spatial dimensions, 
by transforming the string frame action into the Einstein frame and by doing 
the functional variation directly in the Einstein frame variables \cite{ggv2}. 
Clearly the two approaches lead to the same result expressed in eqs. (\ref{be1})--(\ref{be4}) and (\ref{be2})--(\ref{be3}).

In the Einstein frame, the fluid evolution and the dilaton are directly coupled in the conservation equation 
(\ref{b4}). In the particular case of radiation, since the energy--momentum tensor is traceless, the
coupling to the dilaton disappears from Eq. (\ref{b4}).
 
The solution presented in Eqs. (\ref{pot1})--(\ref{dilaton1}) cannot be written in simple analytical terms in the 
Einstein frame, but Eqs. (\ref{be1}) and (\ref{be4}) can be integrated numerically. 
In the Einstein frame description, the 
asymptotics of the solution (\ref{pot1})--(\ref{dilaton1}) are 
\begin{eqnarray}
&& a(\eta) \simeq a_{-} \sqrt{ -\frac{\eta}{2 \eta_0}}, ~~~~~~~~ a_{-} = e^{-\varphi_0/2} \sqrt{\frac{2(\sqrt{3} +1)}{\sqrt{3}}},
\nonumber\\
&& \varphi_{-} = \varphi_0 - \ln{2} - \sqrt{3} \ln{\biggl(\frac{\sqrt{3} +1}{\sqrt{3}}\biggr)} 
- \sqrt{3} \ln{\biggl(- \frac{\eta}{2 \eta_0}\biggr)},
\nonumber\\
&& {\cal H}_{-} = \frac{1}{2\eta}, ~~~~~~~~~~~~~~~\varphi_{-}' = - \frac{\sqrt{3}}{\eta},
\label{solminus}
\end{eqnarray}
 for $\eta \to -\infty$,  and 
\begin{eqnarray}
&& a(\eta) \simeq a_{+} \sqrt{ \frac{\eta}{2 \eta_0}}, ~~~~~~~~~~~~ a_{+} = e^{\vp_0/2} \sqrt{\frac{2(\sqrt{3} -1)}{\sqrt{3}}}
\nonumber\\
&& \varphi_{+} = \varphi_0 - \ln{2} - \sqrt{3} \ln{\biggl(\frac{\sqrt{3} - 1}{\sqrt{3}}\biggr)}
+ \sqrt{3} \ln{\biggl( \frac{\eta}{2 \eta_0}\biggr)},
\nonumber\\
&&{\cal H}_{+} = \frac{1}{2\eta}, ~~~~~~~~~~~~~~~\varphi_{+}' =  \frac{\sqrt{3}}{\eta},
\label{solplus}
\end{eqnarray}
for $\eta \to +\infty$. The branch of the solution denoted by minus 
describes an accelerated contraction, since the first derivative of the scale 
factor is negative while the second  is positive. The branch of the solution denoted 
with plus describes, in the Einstein frame, a decelerated expansion, since the first derivative of 
the scale factor is positive while the  derivative is negative. In both branches the 
dilaton grows and its derivative is always positive-definite (i.e. $\vp'_{\pm} >0$ ) as required 
by the present approach to bouncing solutions.

Recalling that,  in the Einstein frame, 
the  potential term of  Eq. (\ref{pot1}) and its derivative with respect to $\vpb$ are
\begin{equation}
\frac{\partial V}{\partial\vpb} e^{\vp} a^2 = 4 V e^{\varphi} a^2, ~~~~~ V=  - V_0 \frac{e^{-2\vp}}{a^{12}},
\end{equation}
 Eqs. (\ref{be1})--(\ref{be3}) can be numerically integrated across the bounce.

\begin{figure}
\begin{center}
\begin{tabular}{|c|c|}
      \hline
      \hbox{\epsfxsize = 6.9 cm  \epsffile{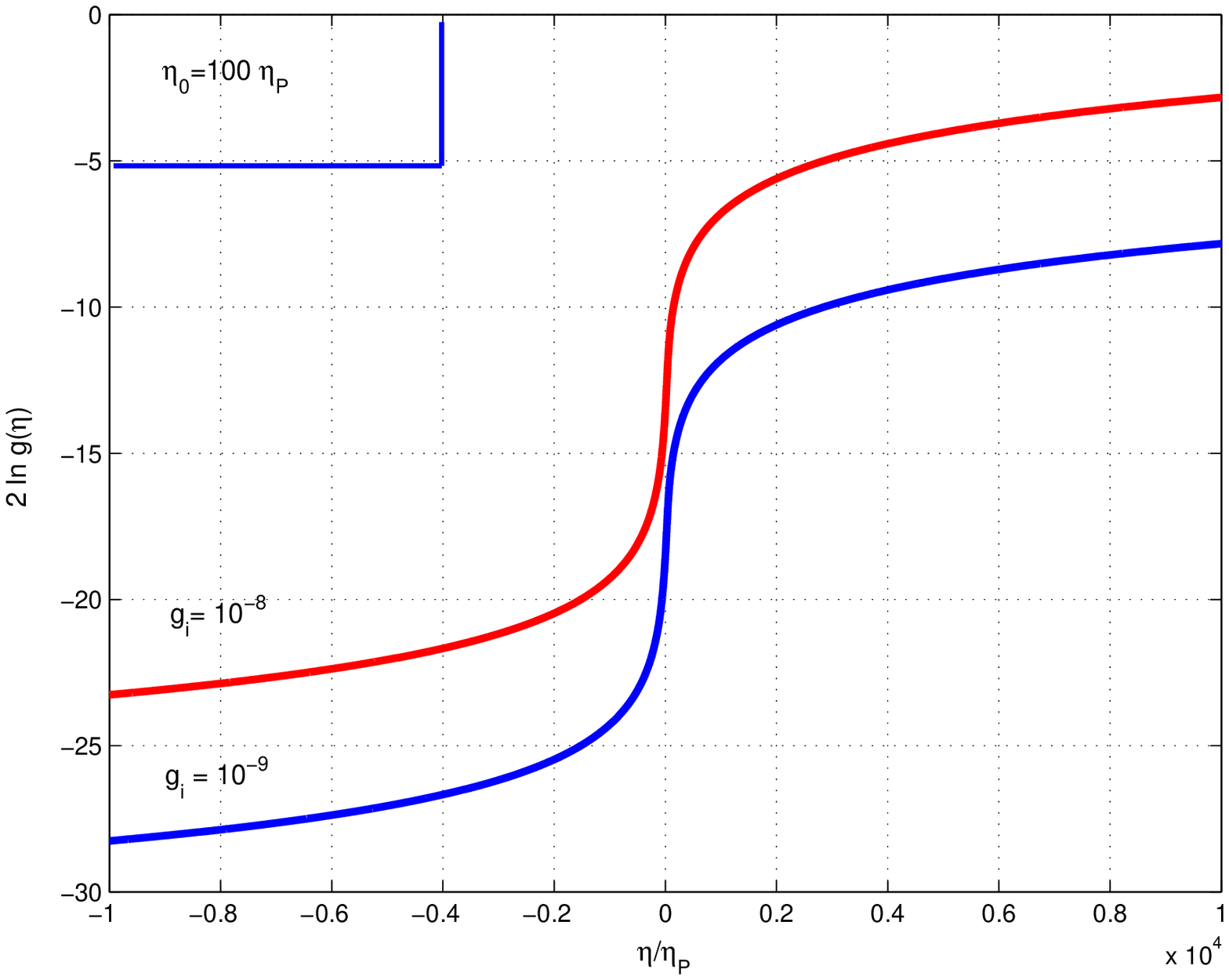}} &
      \hbox{\epsfxsize = 7 cm  \epsffile{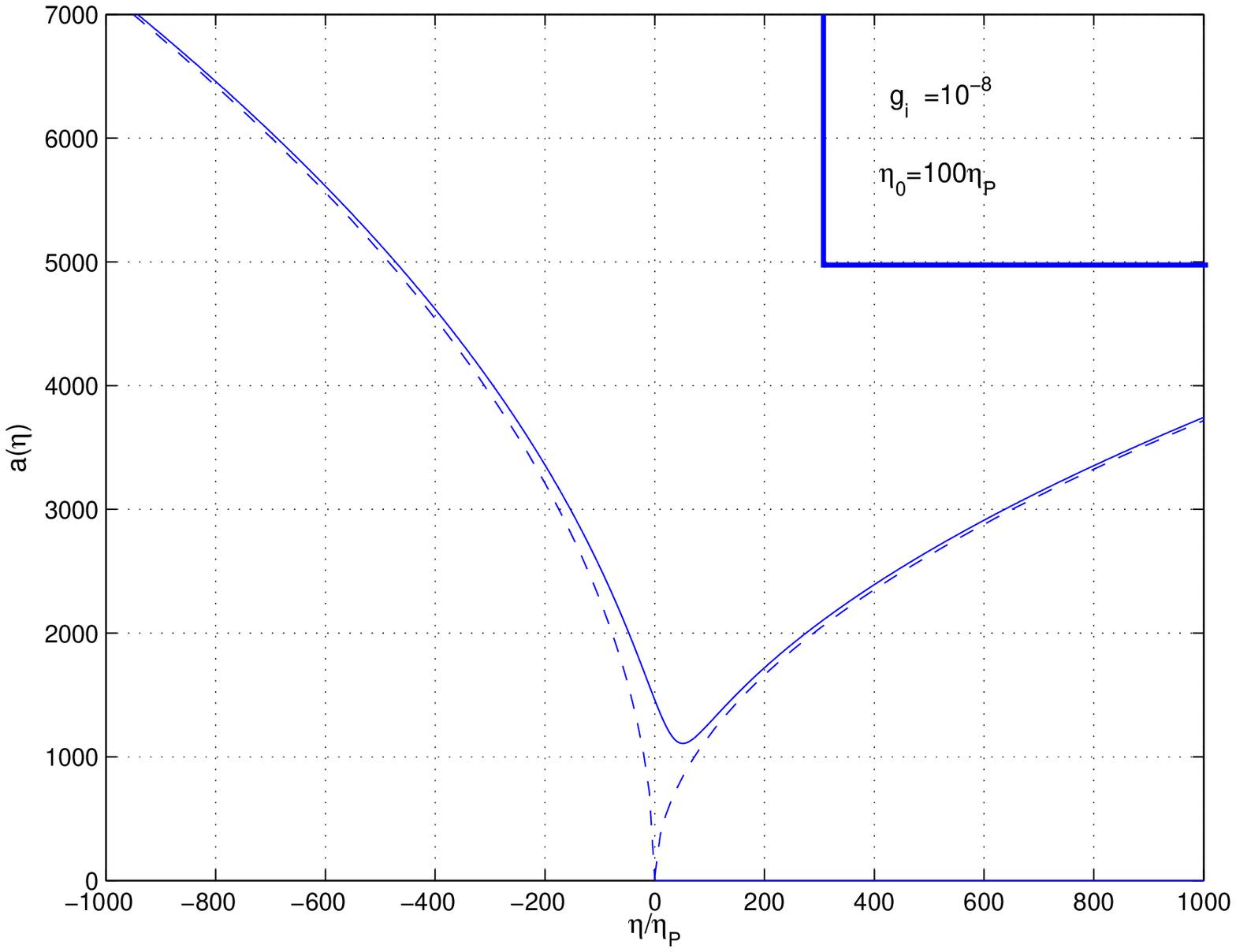}}\\
      \hline
\end{tabular}
\end{center}
\caption{In the left plot the result of the integration is illustrated 
in terms of the (Neperian) logarithm of the gauge coupling.
In the right plot the scale factor obtained numerically (solid line) 
is compared with the analytical (asymptotic) solutions (dashed lines) 
following from Eqs. (\ref{solminus}) and (\ref{solplus}). For both integrations $\eta_{\rm i} = - 10^{7} \eta_{\rm P}$.} 
\label{figure1}
\end{figure}

By setting  initial 
conditions on the asymptotic solution, the integration of the  non-linear problem can be performed by checking that the constraint 
of Eq. (\ref{be1}) is always accurately satisfied. For instance, in the case 
of the numerical solutions illustrated in Fig. \ref{figure1},  
the constraint is satisfied with a precision of $10^{-8}$ over the whole 
range of the solution. 

In  the  light of the subsequent applications, it is useful to parametrize 
the numerical evolution of the solution in terms of the gauge coupling and of its derivatives
\begin{equation}
g(\eta) = e^{\varphi/2},~~~~~~ \varphi'(\eta) = 2 \frac{g'}{g}.
\end{equation}

In order to find  cold-bounce solutions, the system of Eqs. (\ref{be1})--(\ref{be3})
is integrated in the absence of sources. The initial conditions are 
chosen in such a way that the gauge coupling and the curvature scale, in Planck units, are 
always minute. Defining  $\eta_{i}$ as the initial time of integration 
\begin{equation}
g_{i} = g(\eta_{i}) \ll 1, ~~~~~~~\frac{{\cal H}(\eta_{i})}{M_{\rm P}}=\frac{{\cal H}_i}{M_{\rm P}} \ll 1.
\end{equation}
The Hamiltonian constraint of Eq. (\ref{be1}) is enforced on the initial data and 
the initial conditions for the scale factor coincide with the solution (\ref{solminus}).
One important parameter of the solutions is the width of the bounce, which will 
be conventionally denoted by $\eta_0$. The specific value of $\eta_0$ 
for a given solution can be determined numerically. Typical values of $\eta_0$ 
range between $\eta_{\rm P}$ and $100 \eta_{\rm P}$. Furthermore, it turns out 
that, numerically:
\begin{equation}
{\cal H}_{\rm max} \simeq \frac{0.6}{\eta_0}, ~~~~~~\varphi_{\rm max}' = \frac{3.8}{\eta_0}.
\label{est}
\end{equation}
These expessions are rather accurately verified for all the sets of initial conditions
analysed in the present investigation.  
In the following numerical examples the initial time of integration will always be 
denoted by $\eta_{\rm i}$. For each integration the initial value of the gauge coupling, $g_{i}$ 
will also be  specified.

\begin{figure}
\begin{center}
\begin{tabular}{|c|c|}
      \hline
      \hbox{\epsfxsize = 7.3 cm  \epsffile{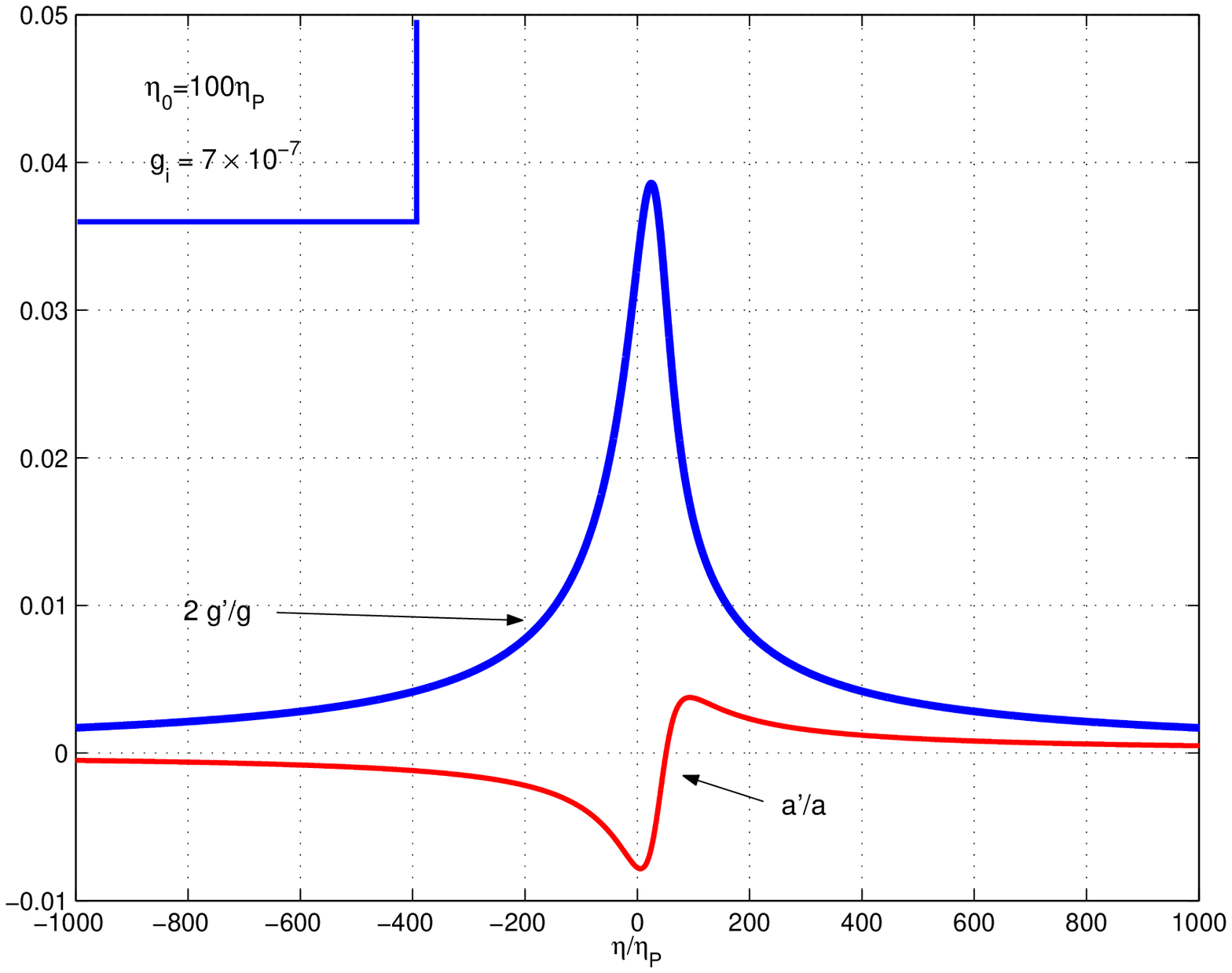}} &
      \hbox{\epsfxsize = 7 cm  \epsffile{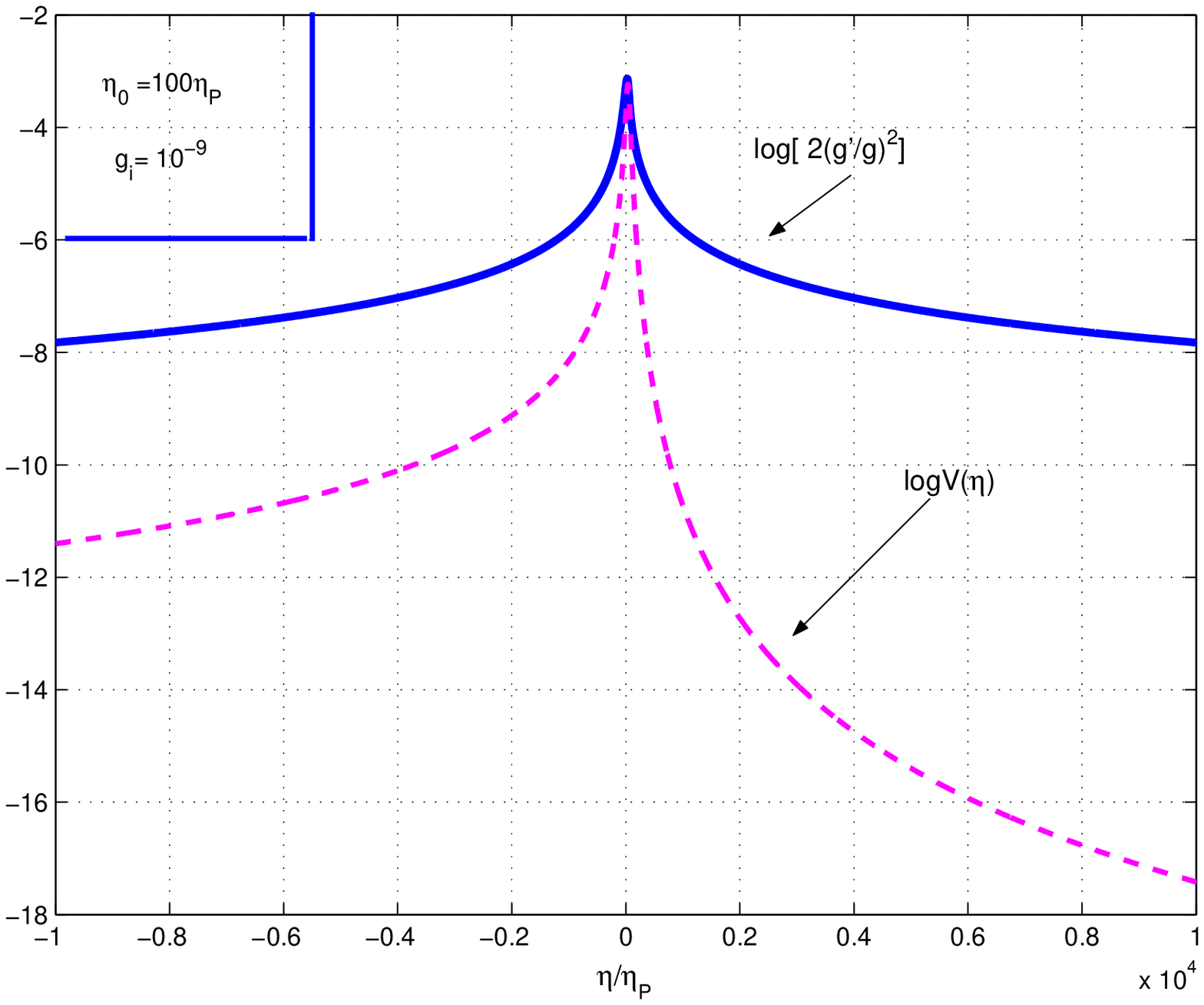}}\\
      \hline
\end{tabular}
\end{center}
\caption{In the left plot cold-bounce solutions are 
illustrated in terms of $a'/a$ and in terms of $g'/g$. In the right plot 
the dilaton potential is compared with the dilaton kinetic term. The various quantities are 
indicated above or below each curve. As in Fig. \ref{figure1}, $\eta_{i}= -10^{-7} \eta_{\rm P}$.}
\label{figure2}
\end{figure}

In Fig. \ref{figure1} the results of the numerical integration 
are illustrated, for different sets of initial conditions in terms of the evolution of the dilaton (i.e. 
twice the Neperian logarithm\footnote{In the present paper the logarithm in ten basis will be denoted by $\log$ while the Neperian 
logarithm will be denoted by $\ln$.} of the gauge coupling) and of the scale factor. In Fig. \ref{figure2} the evolution of the 
derivative of the dilaton and of the Hubble parameter is reported in conformal time.

In the left plot of Fig. \ref{figure1}, the 
numerical solution for the evolution of the scale factor is compared with 
the asymptotic solutions provided by the analytical expressions 
derived in Eqs. (\ref{solminus}) and (\ref{solplus}). The analytical approximation 
reproduces  rather well with the numerical result. The reason for this 
agreement stems from the right plot in Fig. \ref{figure2} where the logarithm (in ten basis) of the  potential terms is 
compared with the logarithm of the  dilaton kinetic energy. Far from the bouncing region the potential 
is always orders of magnitude smaller than the dilaton kinetic energy. Hence the asymptotic solutions are, in practice, a good 
approximation to the full solution not only for $\eta \to \pm \infty$ but also, in less restrictive terms, away from the 
bounce.
The solutions of Fig. \ref{figure1} and of the following plots  will be presented in terms of the ratio $\eta/\eta_{\rm P}$ 
where $\eta_{\rm P} = \sqrt{2}/M_{\rm P}$. In natural Planck units, $ 16\pi G= 2/M_{\rm P}^2=1$ and $ \eta_{\rm P} =1$.

The solutions illustrated so far have realistic features. The most notable feature  being 
the possibility of describing the transition between the pre-big bang and the post-big bang 
at relatively small curvature. 
However,  various difficulties still have to be faced. 
The first problem is that the Universe emerging after the bounce 
is {\em cold}. The solution, even if expanding, is not dominated by radiation 
as one would like it  to be at some later stage of the life of the Universe.
The second problem is connected with the evolution of the gauge coupling. In 
order to trust our perturbative extimates, it should happen that 
$g(\varphi)<1$ throughout all the stages of the model with an aymptotically 
(constant) value of the order of  $0.1$ or $0.01$. In the present case, on the contrary, the 
dilaton coupling is always  growing as a power of the conformal time after the bounce.

\renewcommand{\theequation}{ 3.\arabic{equation}}
\setcounter{equation}{0}
\section{Amplification of Abelian gauge bosons}
In the low-energy string effective action, massless Abelian gauge bosons are 
expected \cite{effac} as a result of the compactification procedure. The modes of the Abelian gauge bosons 
are super-adibatically amplified \cite{ggv3} and they re-enter, at different conformal times, after the bounce.
The modes of comoving frequencies comparable with the typical 
curvature at the bounce  
will be the first ones to re-enter. Furthermore the high-frequency modes 
are also the most energetic and effectively behave like a fluid 
of high-frequency primordial photons.

Right at the bounce the energy density of the primordial  photons 
is still small.   However, as the geometry evolves, the 
energy density of the radiation fluid decreases slower than the 
energy density of the dilatonic sources and, therefore, after a transient time, 
a radiation-dominated phase sets in. The typical transient time 
depends upon the maximal-curvature scale at the bounce and upon the number 
of massless species, which are parametrically amplified. Since in the present case
only one Abelian gauge boson is present, the obtained estimates represent 
the minimal achievable efficiency of the process.

More specifically, consider the case where the coupling of the four-dimensional dilaton field 
to the gauge kinetic term is parametrized as 
\begin{equation}
S_{\rm gb} = -\frac{1}{4} \int d^{4} x \sqrt{- G} e^{- \varphi} F_{\alpha\beta}F^{\alpha\beta}.
\label{gaugeaction}
\end{equation}
Defining the canonically normalized vector potential $A_{i} = g(\eta) {\cal A}_{i}$, 
in the radiation gauge 
\begin{equation}
{\cal A}_{0} =0,~~~~~~~~~\vec{\nabla} \cdot \vec{{\cal A}} =0,
\end{equation}
the canonical Lagrangian density can be obtained from Eq. (\ref{gaugeaction})
by dropping total derivatives: 
\begin{equation}
{\cal L}(\vec{x},\eta) = \frac{1}{2} \biggl[ (\partial_{\eta} {\cal A}_{i})^2  + (g^{-1})'' g {\cal A}_{i}^2  
- (\partial_{i} {\cal A}_{j})^2\biggr],
\end{equation}
The two physical polarizations of the photon can be quantized according to the 
standard rules of quantization in the radiation gauge in curved space-times \cite{ford2}:  
\begin{equation}
\hat{\cal A}_{i}(\vec{x}, \eta) = \sum_{\alpha} \int \frac{d^3 k}{(2\pi)^{3/2}} \biggl[ \hat{a}_{k,\alpha} e^{\alpha}_{i}{\cal A}_{k}(\eta) e^{- i \vec{k}\cdot\vec{x}} +  \hat{a}_{k,\alpha}^{\dagger} e_{i}^{\alpha} {\cal A}_{k}(\eta)^{\star} e^{ i \vec{k}\cdot\vec{x}}\biggr],
\label{ahat}
\end{equation}
and 
\begin{equation}
\hat{\pi}_{i}(\vec{x}, \eta) = \sum_{\alpha} \int \frac{d^3 k}{(2\pi)^{3/2}} \biggl[ \hat{a}_{k,\alpha} e^{\alpha}_{i} \Pi_{k}(\eta) e^{- i \vec{k}\cdot\vec{x}} +  \hat{a}_{k,\alpha}^{\dagger} e_{i}^{\alpha} \Pi_{k}(\eta)^{\star} e^{ i \vec{k}\cdot\vec{x}}\biggr],
\label{pihat}
\end{equation}
where $e_{i}^{\alpha}(k)$ describe the polarizations of the photon and 
\begin{equation}
\Pi_{k}(\eta) = {\cal A}_{k}'(\eta), ~~~~~~~~~ [\hat{a}_{k,\alpha},\hat{a}_{p,\beta}^{\dagger} ] = \delta_{\alpha\beta} 
 \delta^{(3)}(\vec{k} - \vec{p}).
\label{defpi}
\end{equation}

\begin{figure}
\centerline{\epsfxsize = 9cm  \epsffile{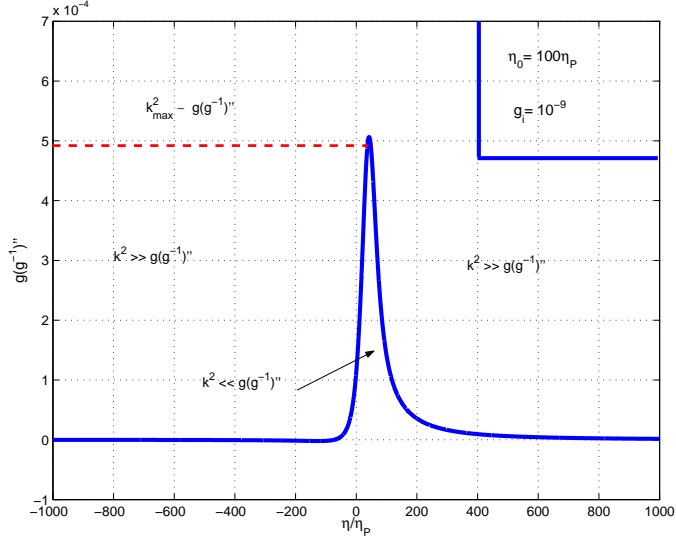}}
\vskip 3mm
\caption[a]{The pump field 
for the primordial  photons (vertical axis) is plotted. For different values of $\eta_0$ the overall 
dimensions of the plot are rescaled (since $g(g^{-1})''\sim \eta_0^{-2}$),
 but the relation determining $k_{\rm max}\eta_0$  always stays the same. In all the plots of the present 
section $\eta_{\rm i} = -10^{7} \eta_{\rm P}$.}
\label{figure3} 
\end{figure}

The evolution equation for the mode functions will then be, in Fourier space,
\begin{equation}
{\cal A}_{i}'' + \biggl[ k^2 - g (g^{-1})''\biggr] {\cal A}_{i} =0,
\label{modef1}
\end{equation}
where, following the terminology of \cite{gr2,gr3}  the ``pump field''\footnote{Notice that the ``pump field'', i.e. the field transferring energy 
from the background to the fluctuations, is determined 
by the first and second derivatives of the gauge coupling.} can  also be  expressed as:
\begin{equation}
g(g^{-1})''= \biggl( \frac{{\vp '}^2}{4} - \frac{\vp ''}{2} \biggr).
\end{equation}
Equation (\ref{modef1}) tells us that all the modes  
$k^2 < |g (g^{-1})''|$ are super-adiabatically amplified. Naively, the maximal
amplified frequency will then be $k_{\rm max}^2= |g (g^{-1})''|$. 
Since, for $\eta\to \pm \infty $,  $g(g^{-1})'' \simeq \eta^{-2}$, modes with $k\eta >1$ are said to be inside the horizon, 
while modes $k\eta <1$ are outside the horizon. This not fully accurate terminology  will also be employed 
in the following only for sake of simplicity. The maximally amplified modes are then the ones for which 
\begin{equation}
k_{\rm max}^2 \simeq |g (g^{-1})''|,  
\end{equation}
as  is illustrated directly in Fig. \ref{figure3} for a particular choice of the parameters.
  
The Fourier  modes appearing in Eq. (\ref{modef1})  have to be normalized while they are 
inside the horizon for large and negative $\eta$. In this limit the initial conditions provided by quantum mechanics are   
\begin{eqnarray}
{\cal A}_{k}(\eta) &=& \frac{1}{\sqrt{2 k}} e^{- i k\eta},
\nonumber\\
\Pi_{k}(\eta) &=& - i \sqrt{\frac{k}{2}} e^{- i k\eta}. 
\label{NORM}
\end{eqnarray}

In the limit $\eta \to +\infty$ the positive and negative frequency modes will be mixed, 
so that the solution will be represented in the plane wave orthonormal basis as 
\begin{eqnarray}
{\cal A}_{k} &=& \frac{1}{\sqrt{2 k}} \biggl[ c_{+}(k) e^{- i k\eta} + c_{-}(k) e^{ i k\eta}\biggr],
\nonumber\\
{\cal A}_{k}' &=& - i \sqrt{\frac{k}{2}}  \biggl[ c_{+}(k) e^{- i k\eta} - c_{-}(k) e^{ i k\eta}\biggr].
\end{eqnarray}
where $c_{\pm}(k)$ are the (constant) mixing coefficients. The following two relations fully determine 
the square modulus of each of the two mixing coefficients in terms of the complex 
wave-functions obeying Eq. (\ref{modef1}): 
\begin{eqnarray}
&& |c_{+}(k)|^2 - |c_{-}(k)|^2 = i( {\cal A}_{k}^{\star} \Pi_{k} -{\cal A}_{k} \Pi_{k}^{\star}) ,
\label{CMIN}\\
&& |c_{+}(k)|^2 + |c_{-}(k)|^2 = \frac{1}{k^2}\biggl(|\Pi_{k}|^2 + k^2|{\cal A}_{k}|^2\biggr).
\label{CPL}
\end{eqnarray}
After having numerically computed the time evolution
of the properly normalized mode functions, Eqs. (\ref{CMIN}) and (\ref{CPL}) can be used to infer 
the value of the relevant mixing coefficient (i.e. $c_{-}(k)$).  Equation (\ref{CMIN}) is, in fact, 
the Wronskian of the solutions. If the second-order differential equation is written in the form 
(\ref{modef1}),  the Wronskian is always conserved throughout the time evolution of the system. 
Since, from Eq. (\ref{NORM}),  the Wronskian  is equal to $1$ initially, it will  be equal to $1$ all along the time evolution.
Thus, from Eq. (\ref{CMIN}) 
$|c_{+}(k)|^2 = |c_{-}(k)|^2 + 1$. The fact that the Wronskian must always be equal to $1$ is the measure 
of the precision of the algorithm. 

\begin{figure}
\begin{center}
\begin{tabular}{|c|c|}
      \hline
      \hbox{\epsfxsize = 7 cm  \epsffile{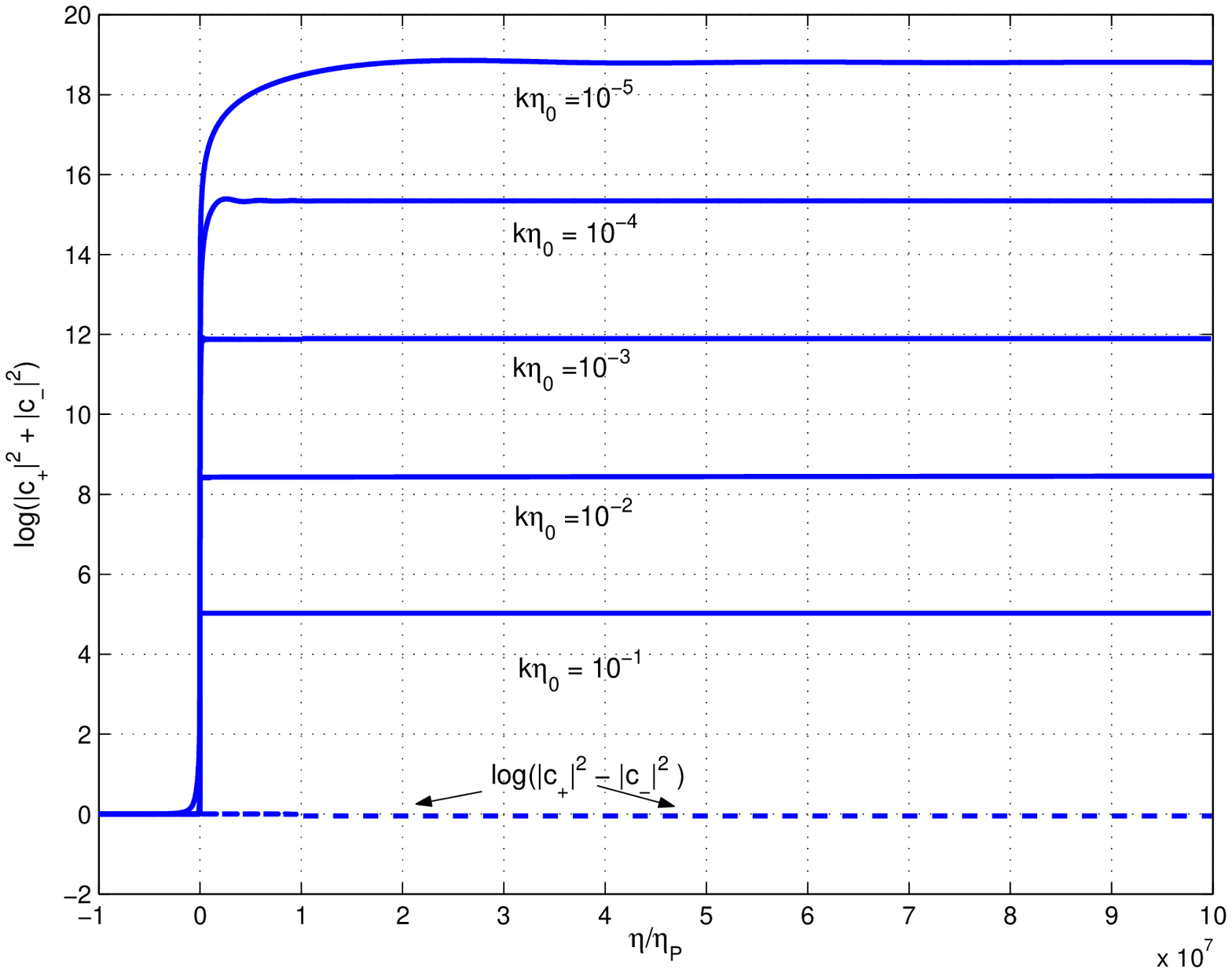}} &
      \hbox{\epsfxsize = 7 cm  \epsffile{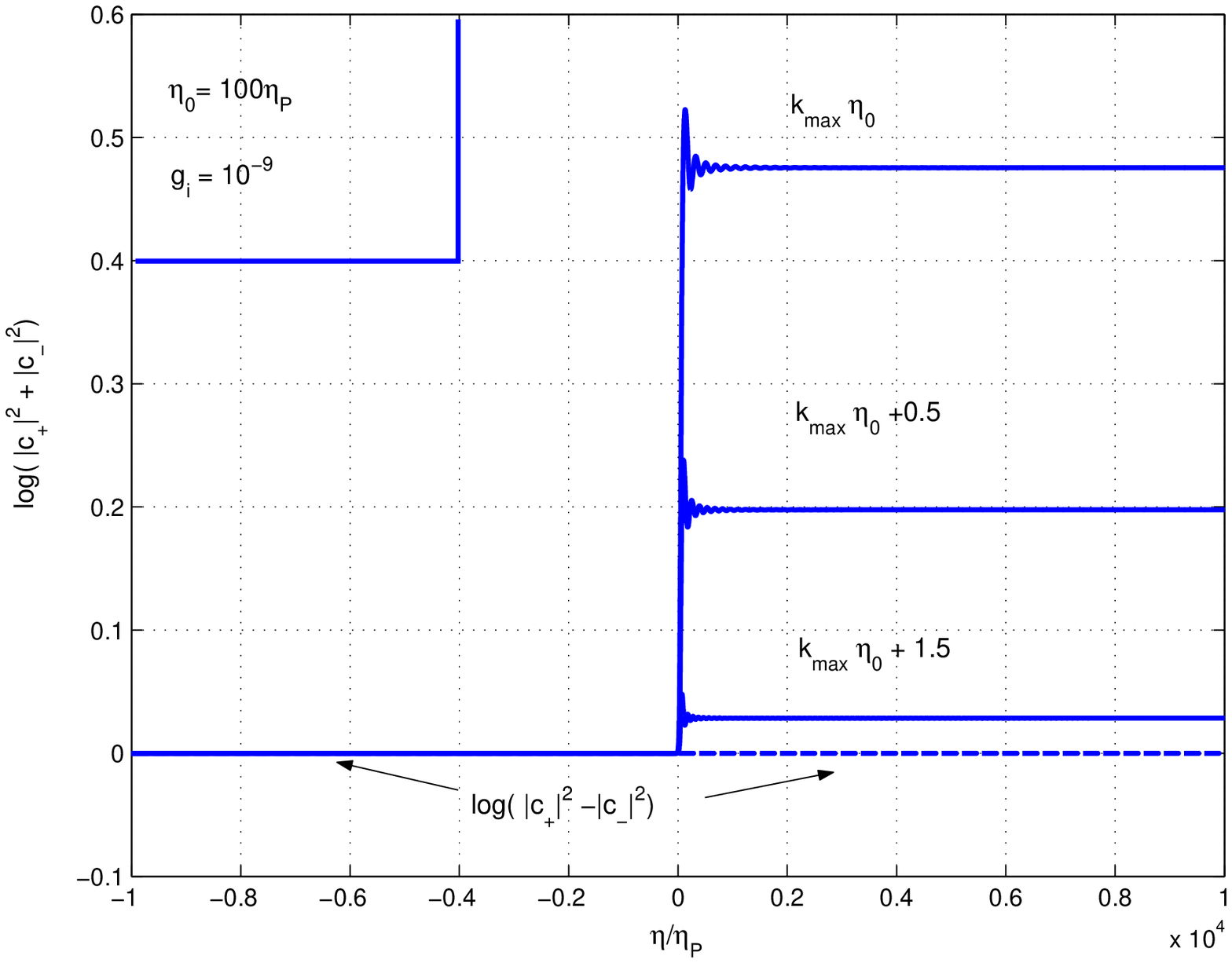}}\\
      \hline
\end{tabular}
\end{center}
\caption{In the left plot the amplification of modes $k^2 \ll |g (g^{-1})''|$ is illustrated. The time evolution 
of the pump field $g(g^{-1})''$ was  described in Fig. \ref{figure3}. In the right plot the largest amplified $k$, i.e. 
$k\simeq k_{\rm max}$, are discussed. The specific value of $k_{\rm max}$ discussed in Eq. (\ref{kmax}) leads to $|c_{-}(k_{\rm max}) \simeq 1$.}
\label{figure4}
\end{figure}

In Fig. \ref{figure4} the numerical calculation of the spectrum is illustrated for different 
values of $k\eta_0$. In the left plot the mixing coefficients 
are reported for modes $k\ll k_{\rm max}$. In the right plot the mixing coefficients are 
reported for modes around $k_{\rm max}$. Clearly, from the left plot a smaller $k$ leads to a larger mixing coefficient 
which means that the spectrum is rather blue. Furthermore by comparing the amplification of different modes 
it is easy to infer that the scaling law is $|c_{+}(k)|^2 + |c_{-}(k)|^2 \propto (k/k_{\rm max})^{- n_{g}}$ , with 
$n_{g} \sim 3.46$, which is in excellent agreement with the analytical determination of the mixing coefficients leading to $n_{g} = 2 \sqrt{3} \sim 3.46$
[see  below, Eq. (\ref{cminanal})].

The second  piece information that  can be drawn from  Fig. \ref{figure4} concerns  $k_{\rm max}$,  whose specific value 
\begin{equation}
k_{\rm max} \simeq \frac{\sqrt{5}-0.5}{\eta_{0}}.
\label{kmax}
\end{equation}
can be determined numerically for different values of $\eta_0$ and 
will be important for an accurate estimate of the back-reaction effects. 

\begin{figure}
\begin{center}
\begin{tabular}{|c|c|}
      \hline
      \hbox{\epsfxsize = 7 cm  \epsffile{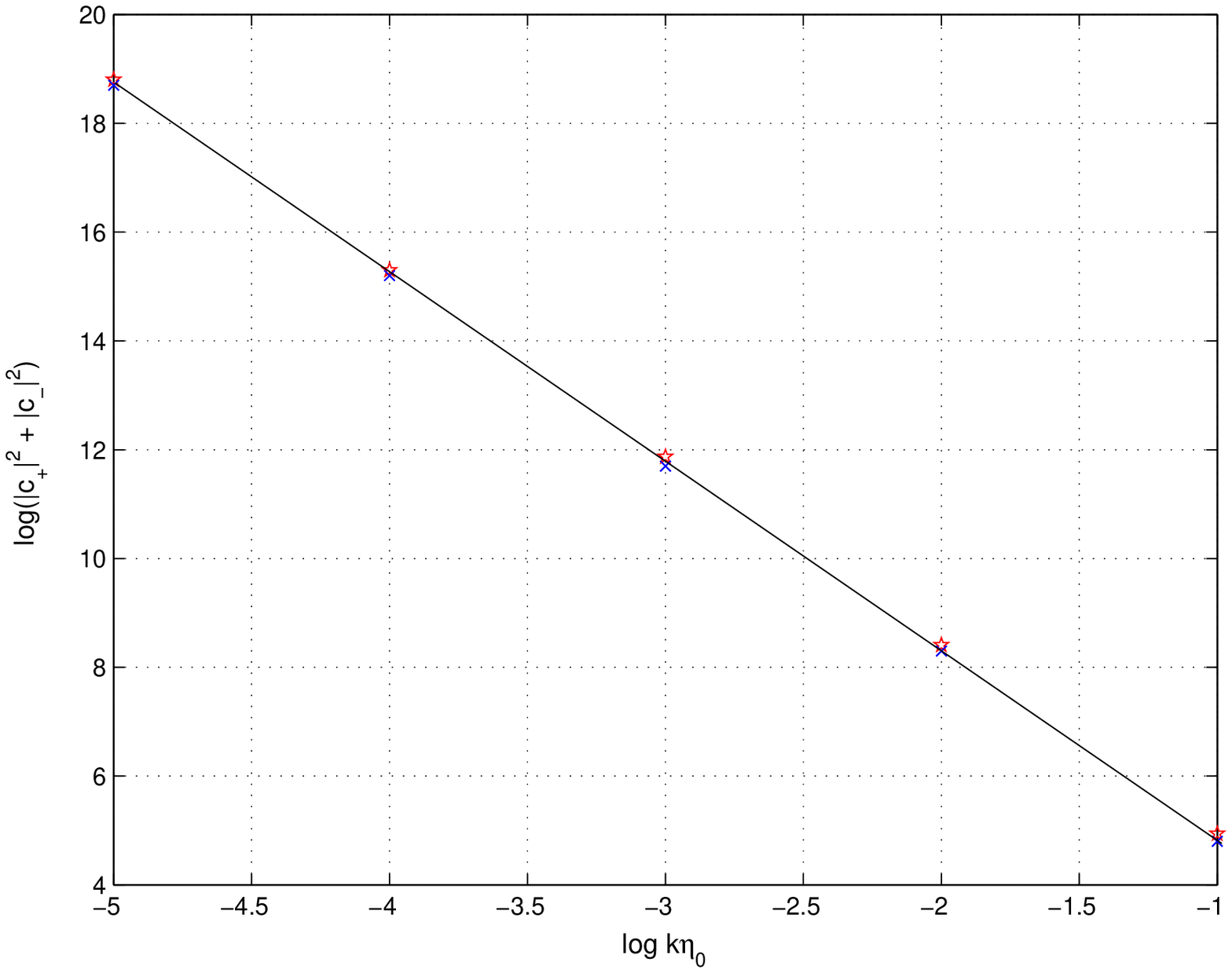}} &
      \hbox{\epsfxsize = 7 cm  \epsffile{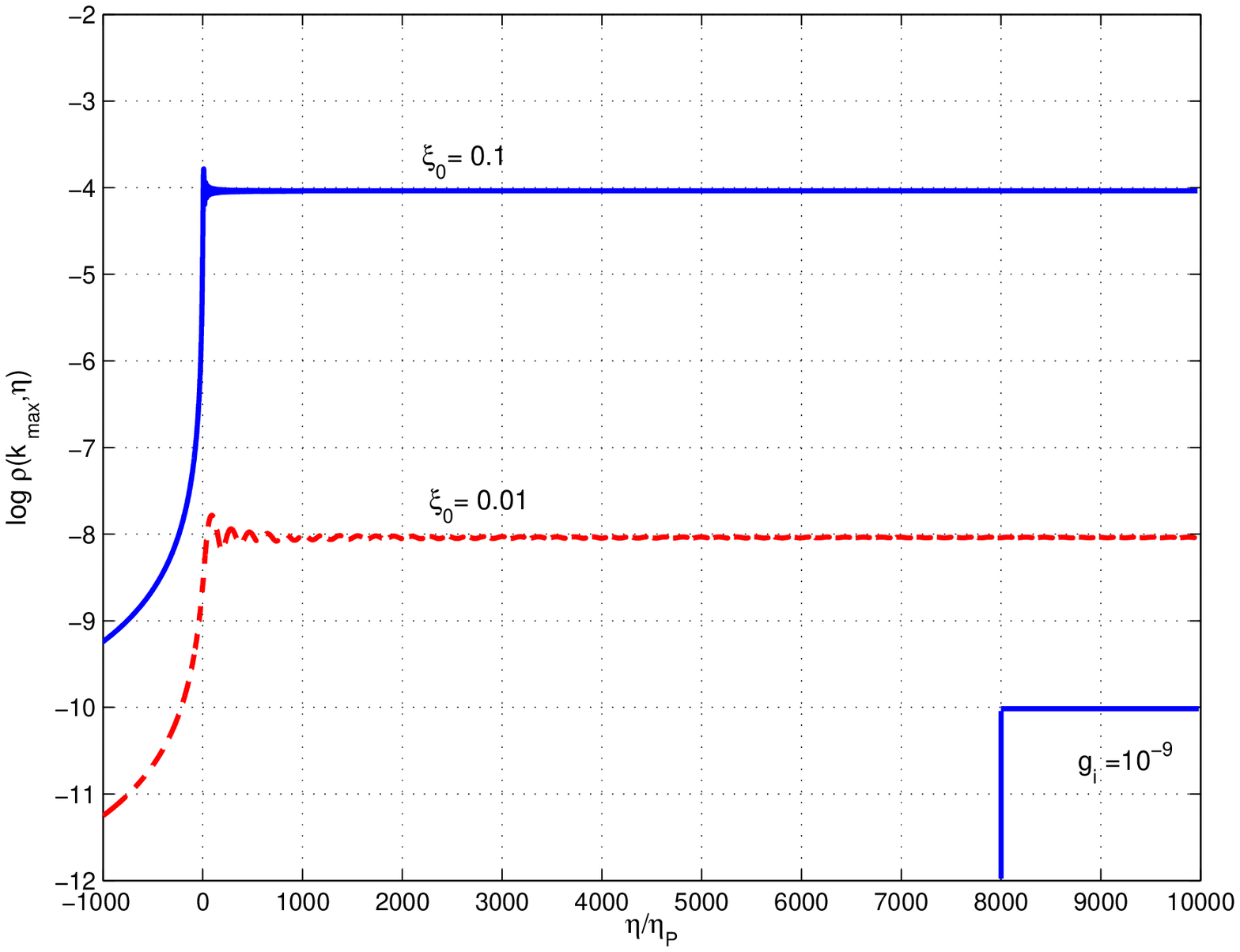}}\\
      \hline
\end{tabular}
\end{center}
\caption{In the left plot the mixing coefficient computed from Eq. (\ref{numest1}) (crosses) are compared with 
the numerical results (stars). The difference is barely visible. In the right plot the logarithmic energy spectrum is 
computed for $k=k_{\rm max}$ and for different sets of parameters. The quantity $\xi_0 = {\cal H}_{0}/M_{\rm P}$ 
has been introduced.}
\label{figure5}
\end{figure}

A reasonably  accurate estimate of the maximal amplified frequency 
is necessary in order to match the analytical estimates with the numerical calculations of the 
back-reaction effects. For the value of $k_{\rm max}$  reported in Eq. (\ref{kmax}), 
the obtained mixing coefficient is $1$, i.e. $|c_{-}(k_{\rm max})| \simeq  1$.
According to Fig. \ref{figure4} (right plot) as we move from $k_{\rm max}$ to larger $k$, $(|c_{+}(k)|^2 + |c_{-}(k)|^2) \simeq
(|c_{+}(k)|^2 - |c_{-}(k)|^2)$ implying that $|c_{-}(k)|  \sim 0$. 
Moreover, from the left plot of Fig. \ref{figure4} it can be appreciated that 
\begin{equation}
|c_{-}(k_{\rm max})|^2 =1, ~~~~~\log{(|c_{+}(k_{\rm max})|^2 +  |c_{-}(k_{\rm max})|^2)} = \log{3} \simeq 0.477.
\end{equation}
Thus the absolute normalization and slope of the relevant mixing coefficient can be 
numerically determined to be 
\begin{equation}
|c_{-}(k)|^2 = \biggl(\frac{k}{k_{\rm max}} \biggr)^{- 2\sqrt{3}}.
\label{numest1}
\end{equation}
In Fig. \ref{figure5} (left plot), the predictions of Eq. (\ref{numest1})  at different $k$ modes 
are compared with the numerical calculation. It can be concluded that Eq. (\ref{numest1}) is rather 
accurate as far as both the slope and the absolute normalization are concerned.

The numerical estimates presented so far can be also corroborated by the usual analytical treatment based on the matching of the 
solutions for the mode functions before and after the bounce. The evolution 
of the modes described by Eq. (\ref{modef1})  can be approximately determined 
from the exact asymptotic solutions given in Eqs. (\ref{solminus}) and (\ref{solplus}),  and  implying that 
$\varphi'_{\pm} \simeq \pm \sqrt{3}/\eta$. Thus the solutions of Eq. (\ref{modef1}) can be obtained in the two asymptotic regimes
\begin{eqnarray}
&& {\cal A}_{k, -}(\eta) = \frac{\sqrt{-\pi\eta}}{2} e^{ i \frac{\pi}{2} ( \nu + 1/2)} H_{\nu}^{(1)}( - k\eta) ,~~~~~~~~~~\eta\leq -\eta_{1},
\nonumber\\
&&  {\cal A}_{k, +}(\eta) = \frac{\sqrt{\pi\eta}}{2} e^{ i \frac{\pi}{2} ( \mu + 1/2)}\biggl[ c_{-} 
H_{\mu}^{(1)}( k\eta) + c_{+} e^{- i \pi (\mu +1/2)} H_{\mu}^{(2)}(k\eta) \biggr] ,~\eta \geq -\eta_{1},
\label{exsol}
\end{eqnarray}
 where $H^{(1,2)}_{\alpha}$ are Hankel functions of first and second kind \cite{magnus} whose related indices are 
\begin{equation}
\nu= \frac{\sqrt{3} -1}{2},~~~~~~~~~~~\mu= \frac{\sqrt{3} +1}{2}.
\label{munu}
\end{equation}
The time scale $\eta_1$ defines the width of the bounce  and, typically, $\eta_{1} \sim \eta_0$.

Phases appearing in Eq. (\ref{exsol}) are carefully chosen so that 
\begin{equation}
\lim_{\eta \to -\infty} {\cal A}_{k} = \frac{1}{\sqrt{2 k}} e^{- i k\eta}.
\end{equation}
Using then the appropriate matching conditions 
\begin{eqnarray}
&& {\cal A}_{k,-}(-\eta_1)= {\cal A}_{k,+}(\eta_1),
\nonumber\\
&& {\cal A}_{k,-}'(-\eta_1)= {\cal A}_{k,+}'(\eta_1),
\end{eqnarray}
and defining $x_1 = k\eta_1$, the obtained mixing coefficients are 
\begin{eqnarray}
&& c_{+}(k) = i \frac{\pi}{4} x_1 e^{i\pi( \nu +\mu + 1)/2} \biggl[ - \frac{\nu + \mu +1}{x_1} H_{\mu}^{(1)}(x_1) H_{\nu}^{(1)}(x_1) 
\nonumber\\
&&+
H_{\mu}^{(1)}(x_1) H_{\nu+ 1}^{(1)} (x_1) + H_{\mu+1}^{(1)}(x_1) H_{\nu}^{(1)}(x_1) \biggr],
\label{cpp}\\
&& c_{-}=  i \frac{\pi}{4} x_1 e^{i\pi( \nu -\mu)/2} \biggl[ - \frac{\nu + \mu +1}{x_1} H_{\mu}^{(2)}(x_1) H_{\nu}^{(1)}(x_1) 
\nonumber\\
&& + H_{\mu}^{(2)}(x_1) H_{\nu+ 1}^{(1)}(x_1) + H_{\mu + 1}^{(2)}(x_1) H_{\nu}^{(1)}(x_1) \biggr],
\label{cpm}
\end{eqnarray}
satisfying the exact Wronskian normalization condition $|c_{+}(k)|^2 - |c_{-}(k)|^2 =1$.
In the small argument limit, i.e. $k\eta_1 \sim k\eta_0 \ll 1$ the leading term in Eq. (\ref{cpm}) leads to 
\begin{equation}
c_{-}(k) \simeq \frac{i~ 2^{\mu +\nu}}{4\pi} e^{i \pi(\nu - \mu)/2} x_1^{- \mu - \nu}  (\nu + \mu -1) \Gamma(\mu) \Gamma(\nu)
\label{cminanal}
\end{equation}
If we now insert the values given in Eq. (\ref{munu}) it turns out that $c_{-}(k) \simeq 0.41~ |k\eta_1|^{- \sqrt{3}}$. The 
spectral slope agrees with the numerical estimate, as already stressed. The absolute normalization cannot be determined
from Eq. (\ref{cminanal}),  where the small argument limit has already been taken. In order to determine the 
absolute normalization the specific value of $k_{\rm max}\eta_1$ has to be 
inserted in Eq. (\ref{cpm}). The result of this procedure, taking $\eta_1\sim \eta_0$ is $|c_{-}(k_{\rm max})|^2 = 0.14$,  which is 
roughly a factor of $10$ smaller than the interpolating formula given in Eq. (\ref{numest1}). 

\renewcommand{\theequation}{4.\arabic{equation}}
\setcounter{equation}{0}
\section{Gravitational reheating of a cold bounce}

A reasonably accurate determination of the energy density of the modes re-entering the horizon after the bounce 
and modifying the dynamics of the background can now be presented.
The expression of the canonical energy--momentum tensor derived from the action (\ref{gaugeaction}) is 
\begin{equation}
T_{\mu}^{\nu} = e^{-\varphi} \biggl[ - F_{\mu\alpha}F^{\nu\alpha} + \frac{\delta_{\mu}^{\nu}}{4} F_{\alpha\beta}F^{\alpha\beta}\biggr].
\label{canenmom}
\end{equation}
From the explicit expression of Eq. (\ref{canenmom}) in terms of the gauge potentials ${\cal A}_{i}$,
it is possible to derive the expectation values of the various components of the canonical energy--momentum tensor. While the derivation of these results 
is included in the Appendix,  the relevant results, for the present calculation, are
\begin{eqnarray}
&&\rho_{\rm r} =\langle T_{0}^{0} (\eta)\rangle = \frac{1}{a^4} \int d\ln{k} \rho(k,\eta) ,
\label{00enmom}\\
&& \rho(k,\eta) =
\frac{k^3}{2\pi^2} \biggl\{ |\Pi_{k}|^2 +\biggl[\biggl( \frac{g'}{g}\biggr)^2 + k^2 \biggr] |{\cal A}_{k}|^2 + \frac{g'}{g}\biggl[ \Pi_{k}^{\ast} {\cal A}_{k} + \Pi_{k}{\cal A}_{k}^{\ast} \biggr]\biggr\}
\label{log}
\end{eqnarray}
and
\begin{equation}
\langle T_{i}^{j}(\eta)\rangle = - \frac{\rho_{\rm r}}{3}\delta_{i}^{j},~~~~~~~~~~~~\langle T_{i}^{0}(\eta)\rangle =0.
\label{ijenmom}
\end{equation}
Sometimes $\rho(k,\eta)$ is called the logarithmic energy spectrum,
since it gives the energy density of the field per logarithmic interval 
of frequency.

For $\eta\to -\infty$ and $|k\eta| >1$ (i.e. when the modes are inside the horizon at the onset of the dynamical evolution) 
the exact mode functions given in Eq. (\ref{exsol})  imply that the logarithmic energy spectrum is 
\begin{equation}
\rho(k,\eta) = \frac{k^4}{4\pi^2} \biggl[ 2 + \frac{3}{k^2 \eta^2} + {\it O}\biggl( \frac{1}{|k\eta|^4}\biggr) \biggr].
\label{log2}
\end{equation}
The first term is nothing but the energy density of the initial vacuum state that can be 
subtracted. The remaining terms are suppressed by powers of $|k\eta|^{-1}$ and are negligible in this regime.

For $|k\eta|<1$ the modes are super-adiabatically amplified. In this case the logarithmic energy density 
becomes,  from Eqs. (\ref{exsol}) and (\ref{log}) 
\begin{equation}
\rho(k,\eta) = \frac{k^4}{8\pi} (-k\eta) \biggl[ |H_{\nu+1}^{(1)}(-k\eta)|^2 + |H_{\nu}^{(1)}(-k\eta)|^2\biggr].
\label{log3a}
\end{equation}
Expanding for $|k\eta| < 1$, and recalling that Eq. (\ref{log3a}) is valid for negative $\eta$, the leading term is 
\begin{equation}
\rho(k,\eta) =  \frac{k^4}{8\pi^3} \Gamma^2(\nu + 1) 2^{2\nu + 2} (-k\eta)^{- 2 \nu -1},
\label{log3}
\end{equation}
where $  2\nu + 1 = \sqrt{3}$. 

Finally, after the bounce and when the modes have re-entered the horizon, the logarithmic energy density 
becomes 
\begin{eqnarray}
\rho(k,\eta) &=&  \frac{k^4}{4\pi^2} \biggl[ 2 |c_{+}(k)|^2 + 2 |c_{-}(k)|^2 
\nonumber\\
&&+ \frac{3}{k^2 \eta^2} ( |c_{+}(k)|^2 +  |c_{-}(k)|^2 +
c_{+}^{\ast}(k) c_{-}(k) e^{ 2 i k\eta} + c_{-}^{\ast} c_{+} e^{- 2 i k\eta} ) 
\nonumber\\
&&+ 
i\frac{\sqrt{3}}{k \eta} ( c_{+}^{\ast}(k) c_{-}(k) e^{2 i k\eta} - c_{-}^{\ast}(k) c_{+}(k) e^{- 2 i k\eta} ) \biggr].
\label{log5}
\end{eqnarray}
This expression also contains subleading terms going as higher powers of $|k\eta|^{-1}$. However, these terms 
are negligible when the modes have already re-entered. Hence, using the relation between the mixing coefficients, the leading 
term appearing in  Eq. (\ref{log5}) can be rewritten as 
\begin{equation}
\rho(k,\eta) \simeq \frac{k^4}{\pi^2} |c_{-}(k)|^2,
\label{log6}
\end{equation}
where the initial zero-point energy has already been subtracted. Notice that this final estimate 
is consistent with the interpretation of $|c_{-}(k)|^2$ as the mean number of produced 
particles per phase-space interval.

With this analytical understanding, we  are then ready to discuss in detail the effect of the produced 
small-scale fluctuations on the dynamics of the background geometry. As expected from the vanishing trace of Eq. (\ref{canenmom}),
 the effective relation 
between the components of the averaged canonical energy--momentum tensor leads to $ 3 p_{\rm r} =\rho_{\rm r}$ 
(see also Appendix B). Thus the effect of $\rho_{\rm r}$ as given in Eq. (\ref{00enmom}) should be inserted back into 
Eqs. (\ref{be1}) and (\ref{be4}) with the result that, for instance, the following equation 
can be obtained:  
\begin{equation}
{\cal H}' = - \frac{2}{3} \biggl(\frac{g'}{g}\biggr)^2 + \frac{5}{6} \frac{V_0}{g^2 a^{10}} - \frac{a^2}{6} \rho_{\rm r}.
\label{backreaction}
\end{equation}
In the numerical code, Eq. (\ref{backreaction}) is consistently 
solved together with the other equations for the dilaton and for the fluctuations of the gauge fields. 
The problem is simplified by the following observation.  The integration over $k$ appearing in Eq. (\ref{00enmom}) 
runs up to $k_{\rm max}$. Since the logarithmic energy spectrum decreases at large-distance scales, the integral 
\begin{equation}
\rho_{\rm r}(\eta) = \frac{1}{a^4(\eta)}\int^{k_{\rm max}} \rho(k,\eta) \frac{d k}{k}
\label{integral}
\end{equation}
is well estimated by the value of the logarithmic energy spectrum at $k_{\rm max}$:
\begin{equation}
\rho_{\rm r}(\eta) = \frac{k_{\rm max}^4}{a^4(\eta)} \epsilon_0 ,
\label{final}
\end{equation}
where $\epsilon_0 = 0.19$.

The result of eq. (\ref{final}) reproduces 
extremely well the numerical results reported, for instance, in Fig. \ref{figure5} (right plot) if we recall the explicit form of $k_{\rm max}$ given in the 
previous section.

The numerical integration   may still be rather stiff 
because of the different scales present in the model (in particular the width of the bounce 
may be  much smaller, for instance, than the typical time at which radiation becomes dominant).
For this purpose the Rosenbrok method (particularly suitable for stiff problems \cite{num}) 
has been used after the bounce.

Before an  illustration of the numerical results, the main semi-analytical expectations 
will be discussed. The typical time scale defining the dominance of radiation can be 
predicted by estimating, numerically, the time of re-entry of $k_{\rm max}$ and 
by comparing the relative balance between the energy density of the background and the 
energy density of the fluctuations. The time of re-entry of $ k_{\rm max}$ is
\begin{equation}
\eta_{\rm max}\sim \frac{1}{k_{\rm max}} \sim \frac{1}{\epsilon_1} \eta_0, 
\end{equation}
where $\epsilon_1\sim  1.73$.

From this expression, assuming that after the bounce the background is still dominated by the kinetic energy of the dilaton, 
the time $\eta_{\ast}$ can be estimated to be
\begin{equation}
\frac{\eta_{\ast}}{\eta_{\rm P}} \simeq \frac{\epsilon_2}{\epsilon_0} \biggl(\frac{H_0}{M_{\rm P}}\biggr)^{-3},
\label{est2}
\end{equation}
where $\epsilon_2 \sim 12.9$.
For instance, for  $H_0/M_{\rm P} \sim 10^{-2}$, and given the indetermination on the numerical 
parameters,  $\eta_{\ast} \simeq 6\times 10^{7} \eta_{\rm P}$.

After $\eta_{\ast}$ the evolution is expected to be dominated by radiation, which means that 
the scale factor should go linearly with the conformal time. During radiation the potential 
term becomes even more negligible than in the case of a cold-bounce 
solution, where it was already negligible (recall Fig. \ref{figure2}, right plot).
Hence, the evolution of the dilaton will be, from the background equation (\ref{be3}),
\begin{equation}
(a^2 \varphi' )' =0
\label{approxdil1}
\end{equation}
which also implies that 
\begin{equation}
\varphi \simeq \varphi_{1} - \varphi_2 \biggl(\frac{\eta_{\rm P}}{\eta} \biggr),
\label{approxdil2}
\end{equation}
where $\varphi_1$ and $\varphi_2$ are integration constants.

A useful function, which can be used as a diagnostic of a fully achieved transition to radiation, 
is 
\begin{equation}
z(\eta) = \frac{a \varphi'}{{\cal H}}.
\label{zeta}
\end{equation}
It should be noticed that $z(\eta) \sim a(\eta)$, if evaluated on the
 cold-bounce solution,  while  $z(\eta)$ is strictly constant , in the case 
of a radiation-dominated background, by 
virtue of Eqs. (\ref{approxdil1}) and 
(\ref{approxdil2}).

\begin{figure}
\begin{center}
\begin{tabular}{|c|c|}
      \hline
      \hbox{\epsfxsize = 7 cm  \epsffile{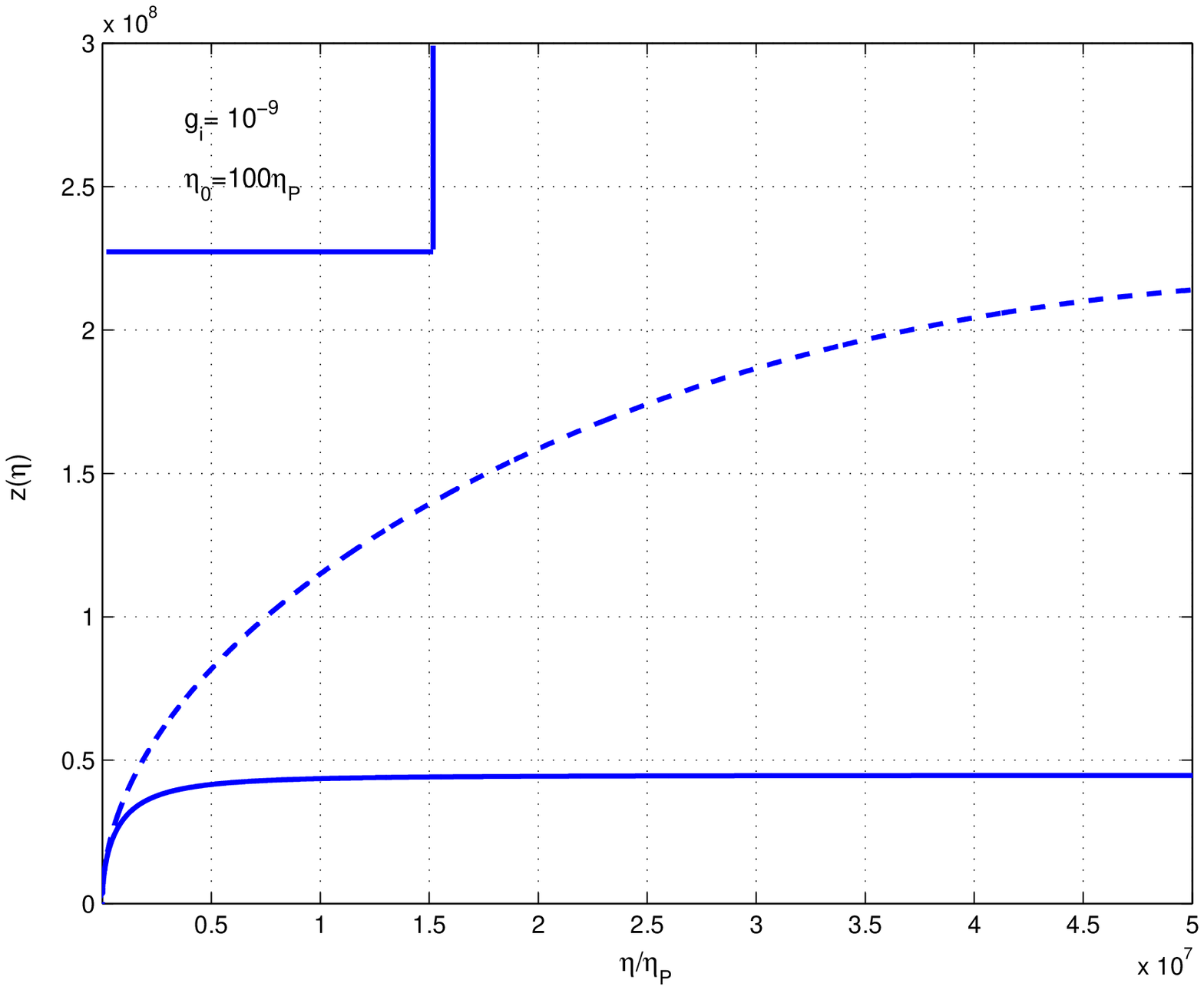}} &
      \hbox{\epsfxsize = 7 cm  \epsffile{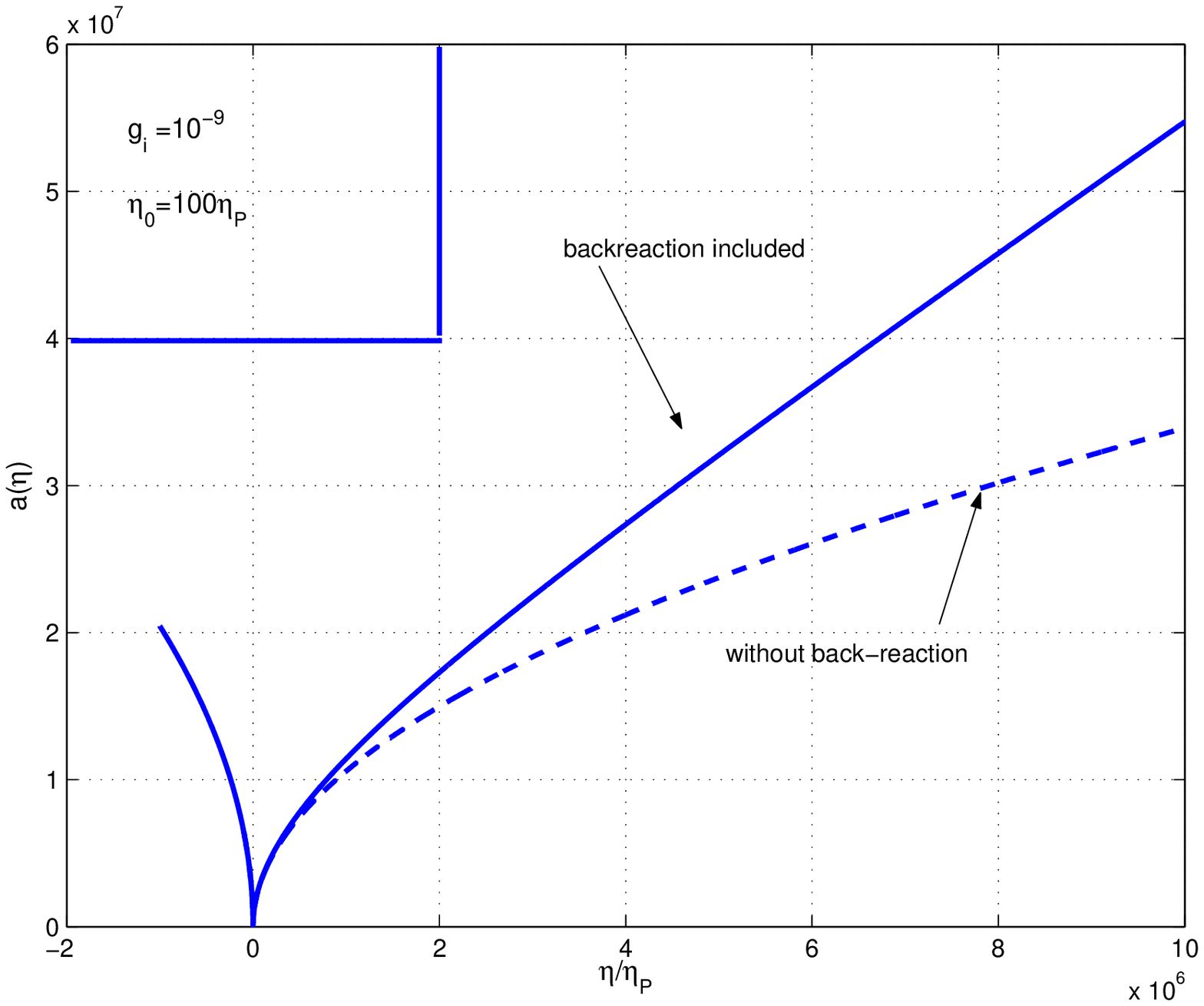}}\\
      \hline
\end{tabular}
\end{center}
\caption{In the left plot the evolution of $z(\eta)$, defined in Eq. (\ref{zeta}), is illustrated. This 
simple function is a diagnostic of a fully achieved transition to radiation where it is constant, while it should grow as $\sqrt{\eta}$ (dotted line) 
in the case when the back-reaction of photons is not included. In the right plot the evolution of the scale factor 
is illustrated with (solid line) and without (dashed line) back-reaction effects.}
\label{figure6}
\end{figure}

\begin{figure}
\begin{center}
\begin{tabular}{|c|c|}
      \hline
      \hbox{\epsfxsize = 7 cm  \epsffile{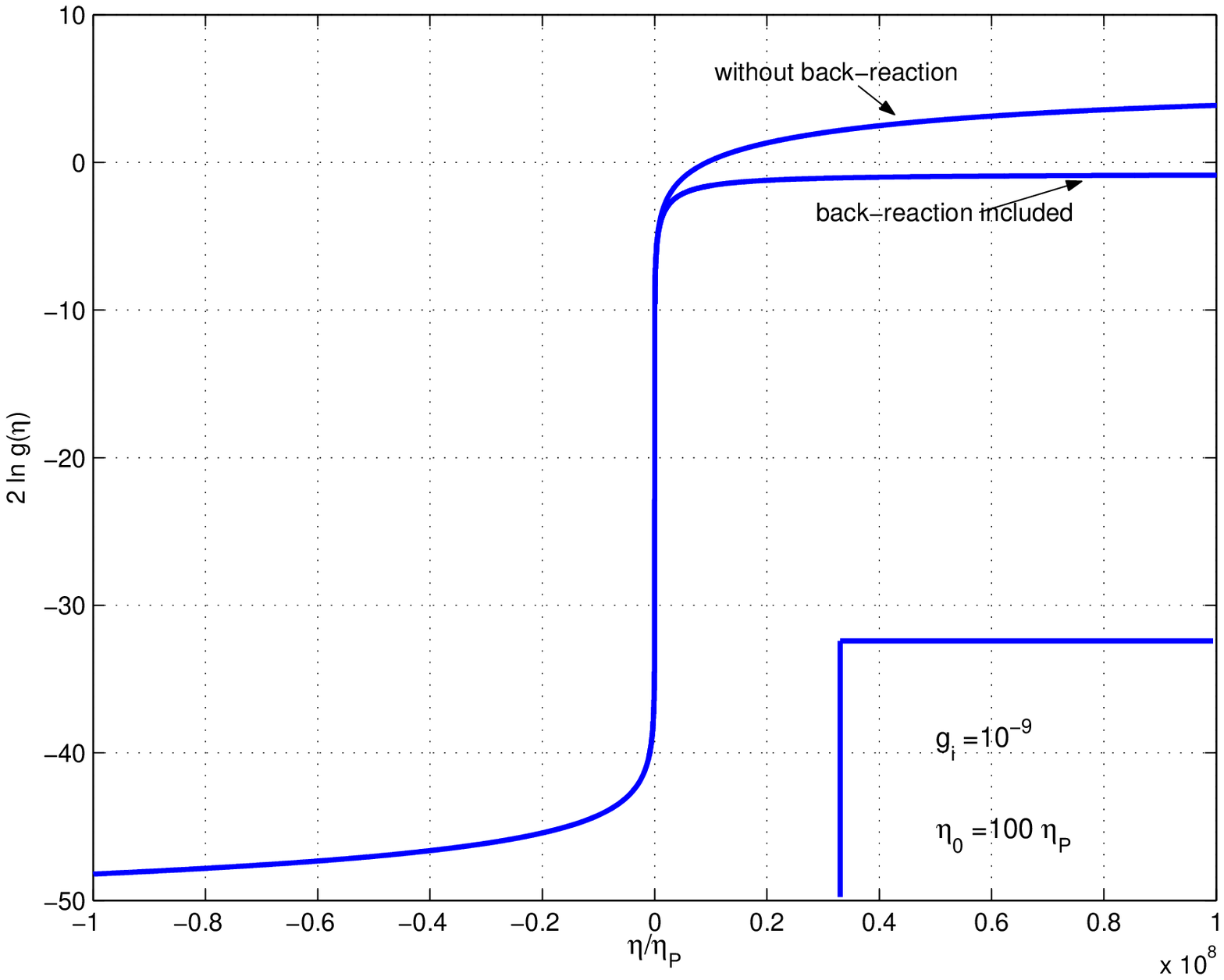}} &
      \hbox{\epsfxsize = 7 cm  \epsffile{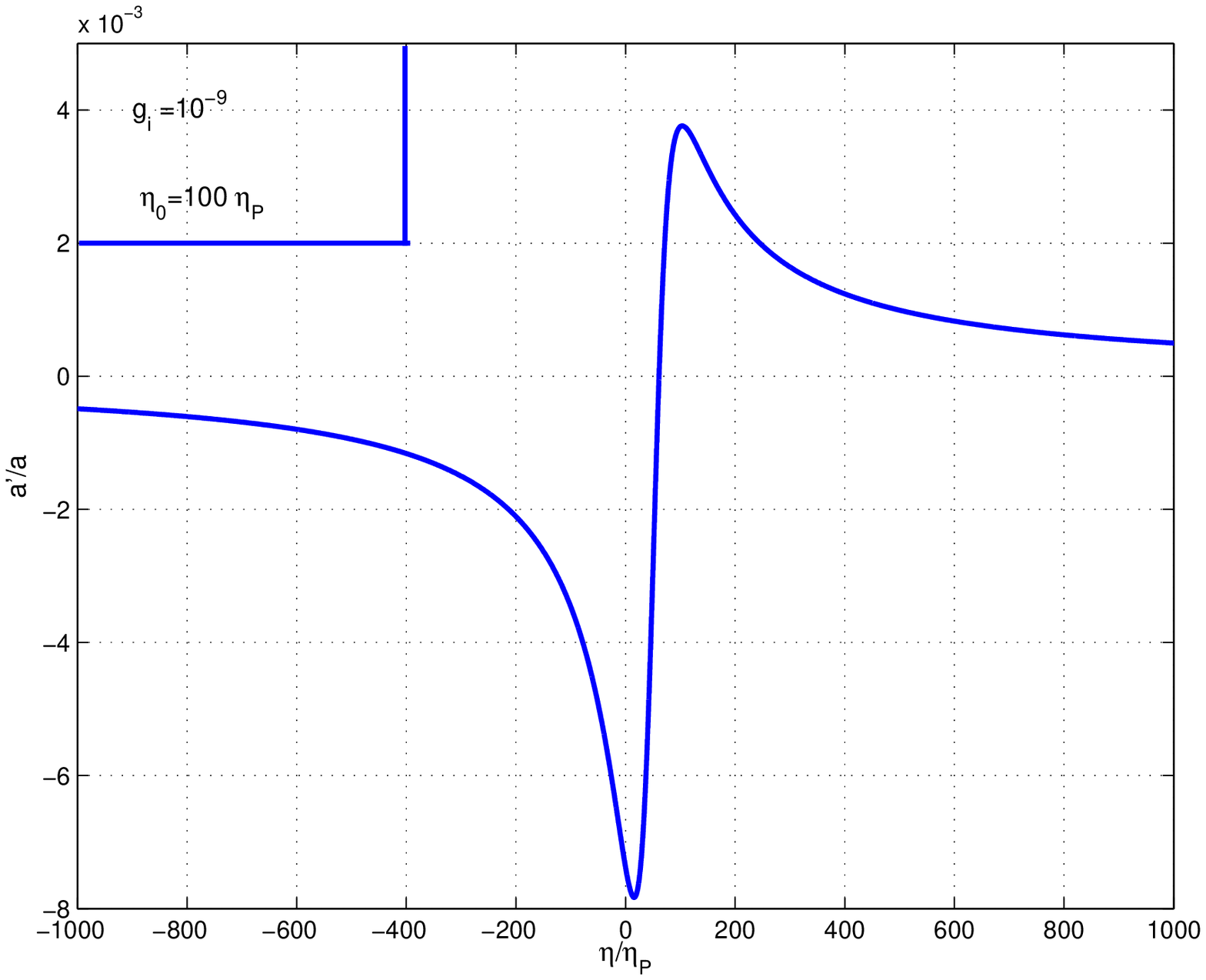}}\\
      \hline
\end{tabular}
\end{center}
\caption{In the left plot the evolution of the logarithm of the gauge coupling is illustrated. 
In the right plot the evolution of $a'/a$ is 
presented.}
\label{figure7}
\end{figure}

The analytical expectations suggested by Eqs. (\ref{est2})--(\ref{zeta}) 
are confirmed by the numerical integration of the full system, which 
is illustrated in Figs. \ref{figure6} and \ref{figure7}.
The constancy of $z(\eta)$ is exactly what is numerically observed. In Fig. \ref{figure6} (left plot), 
after a typical time-scale 
$\eta_{\ast} \sim 10^{6} \eta_{\rm P}$, $z(\eta)$ becomes constant while it should still increase as $z(\eta) \sim a(\eta) \simeq \sqrt{\eta}$
if the  Universe would not be dominated by radiation. At the same time (Fig. \ref{figure6}, right plot) the scale factor grows linearly
and (Fig. \ref{figure7}, left plot) the dilaton goes to constant, as implied by Eq. (\ref{approxdil2}).

\renewcommand{\theequation}{5.\arabic{equation}}
\setcounter{equation}{0}
\section{Concluding discussions}
Cold-bounce solutions have virtues and problems. Their main virtues are that they 
lead to computable models of the bouncing dynamics at low curvatures. 
The main problems are related to the fact that the Universe 
emerging after the bounce is rather cold and dominated by the dilaton.
This occurrence also implies that the dilaton will always be growing, 
before and after the bounce. 

In the present paper a model of gravitational heating of the 
cold bounce has been proposed. The model considers 
only the effect of one species of Abelian gauge bosons,
which are super-adiabatically amplified. The frequencies that are maximally 
amplified are comparable with the typical curvature of the Universe at the bounce
and effectively behave like a fluid of primordial photons. 
The following results have been illustrated: 
\begin{itemize}
\item{} an accurate numerical method for the calculation of the 
amplification of the primordial photons has been developed;
\item{} taking into account the back-reaction of the primordial photons,
the cold-bounce solution can be consistently heated;
\item{} the transition to radiation can occurr for sub-Planckian 
curvature scales;
\item{} solutions have been presented where the dilaton goes to a constant 
value and the asymptotic value of the gauge coupling is always smaller than 
$1$.
\end{itemize}

\section*{Acknowledgements}
The author wishes to thank  G. Veneziano for useful discussions.

\newpage 

\begin{appendix}
\renewcommand{\theequation}{A.\arabic{equation}}
\setcounter{equation}{0}
\section{Solutions with and without fluid sources}
Equations (\ref{b1})--(\ref{b4}) can be written in generic $d$ spatial 
dimensions as:
\begin{eqnarray}
&& \dot{\overline{\varphi}}^2 - d H^2 - V = 
e^{\overline{\varphi}} \overline{\rho},
\label{ba1}\\
&& \dot{H} = \dot{\overline{\varphi}} H + \frac{1}{2} e^{\overline{\varphi}} 
\overline{p},
\label{ba2}\\
&& 2 \ddot{\overline{\varphi}} - \dot{\overline{\varphi}}^2 - d H^2 + V 
- \frac{\partial V}{\partial \overline{\varphi}} =0,
\label{ba3}\\
&& \dot{\overline{\rho}} + d H \overline{p} =0.
\label{ba4}
\end{eqnarray}
In this  Appendix, cold-bounce solutions with and without  fluid 
sources will be presented. It will be shown that 
the naive addition of fluid sources does not stabilize the evolution 
of the dilaton field after the bounce. Since throughout this Appendix we will 
always consider solutions in the string frame, the subscript has been 
omitted in the quantities appearing in Eqs. (\ref{ba1})--(\ref{ba4}).

\subsection{Solutions without fluid sources}
When fluid sources are absent there is a class of regular solutions 
in the presence of exponential potential of the form  
$V(\vpb)= - V_0 e^{\alpha \vpb}$. 
Defining the appropriate variables of the phase space
\begin{equation}
u = \frac{\dot{\overline{\varphi}}}{\sqrt{d} H},~~~~~~~~~~~~~~~~~v = \frac{\sqrt{-V}}{\sqrt{d} H},
\label{uv}
\end{equation}
the constraint (\ref{ba1}) becomes $u^2 + v^2 = 1$ and the equations 
for $u$ and $v$ are 
\begin{eqnarray}
&& \frac{d u}{d\tau} = \sqrt{d} ( 1 - u^2) - \frac{\alpha}{2} \sqrt{d} v^2 \equiv -\frac{\sqrt{d}}{2} (\alpha -2) (1 - u^2),
\label{sourceless0}\\
&& \frac{ d v}{d\tau} = \sqrt{d} \biggl( \frac{\alpha}{2} -1\biggr)u v, 
\label{sourceless1}
\end{eqnarray}
where, in the first equation, the equality follows from $u^2 + v^2 = 1$ and where 
$\tau = \log{a}$. 
 Eqs. (\ref{sourceless0}) and (\ref{sourceless1}) 
 then are in the form of an autonomous system in the plane $(u, v)$.
From Eq. (\ref{uv}) it is clear that
\begin{equation}
t\to - t, ~~~~~~u\to u,~~~~~~~v \to - v.
\label{sym}
\end{equation}
Hence, the  system (\ref{sourceless0})--(\ref{sourceless1}) 
is  symmetric for $u \to u$ and $ v \to -v$ when $\tau \to - \tau$. 
Thanks to the above 
symmetry the phase space of the autonomous system (\ref{sourceless0})--(\ref{sourceless1}) is fully 
characterized by half of the unit disk, for instance, the one for $v\geq 0$.

The critical points of the system  are given by 
\begin{equation}
u_{c} = \pm 1,~~~~~~~~~v_{c} =0.
\label{crit0}
\end{equation}
The plane autonomous system can then  be linearized around the critical points given in 
Eq. (\ref{crit0}), namely 
\begin{eqnarray}
&& u = u_{c} + \epsilon_{ u}, ~~~~v = v_{c} + \epsilon_{v},
\label{deltauv}\\
&& \frac{d\epsilon_{u}}{d \tau} = 2 \mu u_{c} \epsilon_{u},
\label{epsu}\\
&& \frac{d\epsilon_{v}}{d \tau} =  \mu u_{c} \epsilon_{v},
\label{epsv}
\end{eqnarray}
where $2 \mu = \sqrt{d} (\alpha -2)$.
If $u_{c} = 1$ the critical point is an {\em unstable node} since the two 
eigenvalues have the same (positive) sign.
If $u_{c} = -1 $ the critical point is a {\em stable node} 
since the two eigenvalues have the same sign and are both negative.
The unstable node $(1,0)$ corresponds to the accelerated branch of the solution 
where the dilaton coupling increases and the curvature scale also increases.
The stable node $(-1, 0)$ corresponds to the dual (decelerated) solution with 
decreasing dilaton and decreasing curvature.

In this parametrization the evolution equations
can be easily integrated. From Eq. 
(\ref{sourceless0}), integrating 
once and inverting the obtained expression:
\begin{equation}
u(\tau) = \frac{ 1 + {\cal C} e^{2 \mu\tau}}{ 1 - {\cal C} e^{2 \mu\tau}},
\label{utau}
\end{equation}
where the integration constant ${\cal C}$ has to be selected such 
that the appropriate boundary conditions are satisfied. 
The constant  ${\cal C}$ can be chosen in such a way that $u(\tau)$ is everywhere
regular in its domain of definition and, in particular, for $\tau \to 0$.
With this in mind, ${\cal C} =-1$ in order to prevent a possible singularity 
in $\tau =0$. From Eqs. (\ref{sourceless1}) and (\ref{utau}), the 
full result is 
\begin{eqnarray}
&& u(\tau) = - \tanh{ \mu \tau},
\label{us}\\
&& v(\tau) = \frac{1}{\cosh{\mu\tau}},
\label{vs}
\end{eqnarray}
where the possible further integration constant is fixed by the constraint $u^2 + v^2 =1$. 

The solution given by Eqs. (\ref{us}) and (\ref{vs}) interpolates 
between the unstable node $(1, 0)$ and the stable node $(-1, 0)$,  as  is 
apparent from the asymptotic behaviour of $(u,v)$ for $\tau \to \pm \infty$.
Recalling now the explicit definition of $u$ and $v$ in terms of the original 
background fields, i.e. Eqs. (\ref{sourceless0}) and (\ref{sourceless1}), and recalling the explicit 
parametrization of the exponential potential, the following chain of
identities can be obtained:
\begin{eqnarray}
&& u(\tau) = \frac{\dot{\vpb}}{\sqrt{d} H} \equiv \frac{1}{\sqrt{d}} \frac{ d \vpb}{d\tau} = - \tanh{\mu\tau},
\label{chain1}\\
&& v(\tau) = \frac{ \sqrt{ V_0} e^{\frac{\alpha}{2}\vpb}}{\sqrt{d} H}= \frac{1}{\cosh{\mu\tau}}.
\label{chain2}
\end{eqnarray}
Integrating once Eq. (\ref{chain1}) and substituting in (\ref{chain2}):
\begin{eqnarray}
&& H(\tau) = \frac{\sqrt{V_0}}{\sqrt{d}} \frac{ e^{\frac{\alpha}{2} \vpb}}{(\cosh{\mu\tau})^{2/(\alpha -2)}},
\label{H}\\
&& \vpb(\tau) = \vp_{0} - \frac{2}{\alpha -2} \ln{\cosh(\mu\tau)}.
\label{vphi}
\end{eqnarray}
The parametrization in terms of $\tau = \ln{a}$ is then more useful than the one in terms of $t$. 
In fact, Eqs. (\ref{H}) and (\ref{vphi}) have simple analytical expressions in terms 
of $\tau$,  but not necessarily in terms of $t$. 
Recalling that, by definition, $d\tau /dt = H(\tau)$, the explicit relation between $\tau$ and $t$ 
can be worked out by integrating once $H(\tau)$ from Eq. (\ref{H}). 
For instance, in the case $\alpha =4$
\begin{eqnarray}
&& \overline{\varphi} = \varphi_0 - \ln{\cosh{\sqrt{3} \tau}},
\nonumber\\
&& \frac{d \tau}{d t} = \frac{e^{2 \varphi_0} \sqrt{V_0}}{\sqrt{3} \cosh{\sqrt{3} \tau}},
\end{eqnarray}
defining $t_0^{-1}= e^{2 \varphi_0} \sqrt{V_0}$, 
\begin{equation}
t/t_0 = \sinh{\sqrt{3} \tau},
\end{equation}
i.e. $ \tau = \ln{a} = (1/\sqrt{3}) \ln{[ (t/t_0)^2 + \sqrt{(t/t_0)^2 + 1}]}$. This last solution is the one quoted in Eqs. (\ref{pot1})--(\ref{dilaton1}).
Note that if we do not want $H(\tau)$ blowing up for $|\tau|\to \infty$, we have to demand $\alpha \geq 2$. 
In the case $\alpha =2$, $\mu =0$ and Eqs. (\ref{sourceless0})--(\ref{sourceless1}) imply 
that $u$ and $v$ are constant.

\subsection{Solutions with fluid sources}

If fluid sources are included, various solutions are possible. 
Consider first the case $p =0$. In this case Eq. (\ref{b2}) 
can be immediately integrated and an exact regular solution 
of the remaining equations is 
\begin{equation}
V = - V_{0} e^{\vpb} - V_1 e^{4 \vpb},
\label{nai}
\end{equation}
with
\begin{eqnarray}
&& H= \frac{1}{\sqrt{d}}\frac{t_0}{\sqrt{t^2 + t_{0}^2}},
\nonumber\\
&& \vpb = \vp_0 - \frac{1}{2} \log{[ \frac{t^2}{t_0^2} + 1]},
\nonumber\\
&& \rho = \overline{\rho}_0 a^{- d},
\label{nai2}
\end{eqnarray}
subject to the conditions: 
\begin{eqnarray}
&& \overline{\rho}_0 = V_0 ,
\nonumber\\
&& V_1 e^{2 \vp_0} = \frac{1}{t_0^2}.
\end{eqnarray} 

Consider now the case where the barotropic index 
\begin{equation}
w = \frac{p}{\rho} = \frac{\overline{p}}{\overline{\rho}}
\end{equation}
is constant,  
and the dilaton potential is $V = - V_0 e^{2\vpb}$.
In this case, Eqs. (\ref{ba2}) and (\ref{ba3}) can 
be integrated :
\begin{eqnarray}
&& H = w \frac{e^{\vpb}}{ L} ( x + x_1),
\nonumber\\
&& \dot{\vpb} = - \frac{e^{\vpb}}{ L} ( x + x_0),
\label{firstsol}
\end{eqnarray}
where 
\begin{equation}
\frac{ d x}{d t} = \frac{L}{2} \overline{\rho},
\end{equation}
and $ x_1$, $x_0$ are integration constants
\footnote{Since 
$\overline{\rho}$ has dimensions of an inverse length, $x$ and $t$ have the same dimensions.}.

From Eq. (\ref{firstsol}) it can be checked that $\dot{\vp}= \dot{\vpb} + 3 H $ 
is always positive only in the case $w=1/3$, corresponding 
to the case of a radiation fluid. In this case, an appropriate 
choice of the parameters $x_1$ and $x_0$ allows 
a simple analytical solution, which 
can be expressed directly in terms of $x$, i.e. the new time 
variable. If $w = 1/3$, then Eq. (\ref{firstsol}) can be inserted back into Eq. 
(\ref{ba1}), with the result 
\begin{equation}
e^{\vpb} = \frac{3 \overline{\rho} L^2}{3 ( x + x_0)^2 - ( x + x_1)^2 + 3 V_0 L^2}.
\end{equation}
Now we can choose constants to further simplify the general solution.
Choose, in particular,
\begin{equation}
x_0 = L,~~~~~x_1 = 3 x_0,~~~~~~V_0 L^2 = \frac{8}{3}L^2.
\end{equation}
Then, it can be checked that the explicit solution is, in this 
case 
\begin{eqnarray}
&& a(x) = \sqrt{x^2 + L^2} e^{ 3 \arctan{(x/L)}},
\nonumber\\
&& \frac{a'}{a} = \frac{x + 3 L}{x^2 + L^2}, 
\nonumber\\
&& \vpb' = - \frac{ 3 ( x + L)}{ x^2 + L^2}, 
\nonumber\\
&& \frac{d x}{d t} = \frac{\rho_0 L}{2} \frac{1}{a(x)}.
\label{model}
\end{eqnarray}
The shifted dilaton is 
\begin{equation}
\vpb = \vp_0 - 3 \arctan{(x/L)} - \frac{3}{2} \ln{(x^2 + L^2)}.
\end{equation}
Looking at the constraint, we should also bear in mind that
\begin{equation}
e^{\vp_0} = \frac{3}{2} \overline{\rho}_{0} L^2.
\end{equation}

There is a third class of possible solutions of the system of 
Eqs. (\ref{b1})--(\ref{b4}). This class corresponds to the case when 
the barotropic index is a function of time. In this case, Eq. (\ref{firstsol})
is modified, i.e.
\begin{eqnarray}
&& H =  \frac{e^{\vpb}}{ L} \int dx w(x),
\nonumber\\
&& \dot{\vpb} = - \frac{e^{\vpb}}{ L} ( x + x_0).
\label{firstsol2}
\end{eqnarray}
A particularly simple example of this class of solutions is provided 
by the profile \cite{gv1}:
\begin{equation}
w(x) = \frac{1}{d} \frac{x}{\sqrt{x^2 + x_1^2}}.
\label{profile}
\end{equation}
By integrating the equations of the background with a method similar to the 
one introduced in the case of constant $w$ solutions, analytical solutions 
can be found in the case of $V = - V_0 e^{2 \vpb}$. Notice that 
in the case described by Eq. (\ref{profile}) 
the barotropic index interpolates between $w = -1/d$ and $w=1/d$, i.e. 
for $d=3$, $w= -1/3$ and $w = 1/3$. 
The full solution is then:
\begin{eqnarray}
&&\vpb = \vp_0 - \frac{d}{d-1} \log{\biggl[ \biggl(\frac{x}{x_1}\biggr)^2 +1 \biggr]},
\nonumber\\
&& H = \frac{e^{\vp_0}}{d} \biggl[ \biggl(\frac{x}{x_1}\biggr)^2 +1
 \biggr]^{ - \frac{d+1}{2(d-1)}},
\nonumber\\
&& \overline{\rho} = \frac{d-1}{d} e^{\varphi_0} \biggl[ 
\biggl(\frac{x}{x_1}\biggr)^2 +1 \biggr]^{-\frac{1}{d-1}}.
\label{wtdep}
\end{eqnarray}

\renewcommand{\theequation}{B.\arabic{equation}}
\setcounter{equation}{0}
\section{Averages of the energy-momentum tensor}

In this Appendix the averages of the canonical energy--momentum tensor given in Eq. (\ref{canenmom}) will be computed.
In the radiation gauge the components of the energy--momentum tensor read
\begin{eqnarray}
&& T_0^{0} =\frac{1}{2 a^4} \biggl[ ({\cal A}_{i}')^2 + \biggl(\frac{g'}{g}\biggr)^2 {\cal A}_{i}^2 
+ 2 \frac{g'}{g} {\cal A}_{i}' {\cal A}^{i}
+( \partial_{i} {\cal A}_{j})^2 - \partial_{i} {\cal A}_{j} \partial^{j} {\cal A}^{i} \biggr],
\label{001}\\
&& T_{i}^{j} =\frac{1}{a^4} \biggl\{ \partial_{\eta} {\cal A}_{i} \partial_{\eta} {\cal A}^{j} +2 \biggl(\frac{g'}{g}\biggr)^2 {\cal A}_{i} 
{\cal A}^{j} + 2 \frac{g'}{g} \biggl[\partial_{\eta} {\cal A}_{i} \partial_{\eta } {\cal A}^{j} \biggr]
\nonumber\\
&-& 2 \partial_{i} {\cal A}_{k} \partial^{j} {\cal A}^{k} + 2 \partial_{k} {\cal A}_{i} \partial^{j} {\cal A}^{k} 
+\frac{\delta_{i}^{j}}{2} \biggl[ -\partial_{\eta} {\cal A}_{i} \partial_{\eta}{\cal A}^{i} 
\nonumber\\
&-& 2\frac{g'}{g} 
\partial_{\eta} {\cal A}_{i} {\cal A}^{i} - \biggl(\frac{g'}{g}\biggr)^2 {\cal A}_{i} {\cal A}^{i} + 
\partial_{i} {\cal A}_{k} \partial^{i} {\cal A}^{k} - \partial_{i} {\cal A}_{k} \partial^{k} {\cal A}^{i}\biggr]\biggr\}.
\label{ij1}\\
&& T_{i}^{0} = \frac{1}{a^4} \biggl[ \biggl( \partial_{\eta} {\cal A}^{k} + \frac{g'}{g} {\cal A}^{k}\biggr) \biggl( \partial_{i} {\cal A}_{k}
-\partial_{k} {\cal A}_{i} \biggr) \biggr].
\label{0i1}
\end{eqnarray}
Note that in Eqs. (\ref{001})--(\ref{0i1}) the spatial indices are flat. Furthermore, in Eqs. (\ref{ij1}) and (\ref{0i1}) 
the derivation with respect to $\eta$ has been indicated with $\partial_{\eta}$ (and not, as usual, with a prime) 
only in order to avoid confusion with controvariant (flat space) indices.

Recalling now the expression of the field operators given in Eqs. (\ref{ahat}) and (\ref{pihat}), the various 
expectation values implied by Eqs. (\ref{001})--(\ref{0i1}) can be evaluated
separately:  
\begin{eqnarray}
&&\langle \hat{\cal A}_{i}(\vec{x},\eta) \hat{\cal A}_{j}(\vec{y},\eta) \rangle = \int \frac{d^{3} k}{(2\pi)^3} P_{ij}(k) |{\cal A}_{k}(\eta)|^2 
e^{- i \vec{k}\cdot(\vec{x} - \vec{y}) },
\nonumber\\
&& \langle \Pi_{i}(\vec{x},\eta) \Pi_{j}(\vec{y},\eta) \rangle = \int \frac{d^{3} k}{(2\pi)^3} P_{ij}(k) |\Pi_{k}(\eta)|^2 
e^{- i \vec{k}\cdot(\vec{x} - \vec{y}) },
\nonumber\\
&&= \int \frac{d^{3} k}{(2\pi)^3} P_{ij}(k) [\Pi_{k}(\eta)^{\star} {\cal A}_{k}(\eta) + \Pi_{k}(\eta) {\cal A}_{k}(\eta)^{\star} ]
e^{- i \vec{k}\cdot(\vec{x} - \vec{y}) },
\label{expect}
\end{eqnarray}
where 
\begin{equation}
P_{ij}(k) = \sum_{\alpha} e^{\alpha}_{i} e^{\alpha}_{j} = \delta_{i j} - \frac{k_{i} k_{j}}{k^2}.
\end{equation}
Using Eqs. (\ref{expect}) (for spatially coincident points)  into Eqs. (\ref{001}) we obtain
\begin{equation}
\langle T_{0}^{0} (\eta)\rangle = \frac{1}{a^4} \int \frac{d^3 k}{(2\pi)^3}
 \biggl\{ |\Pi_{k}|^2 +\biggl[\biggl( \frac{g'}{g}\biggr)^2 + k^2 \biggr] |{\cal A}_{k}|^2 + \frac{g'}{g}\biggl[ \Pi_{k}^{\ast} {\cal A}_{k} + \Pi_{k}{\cal A}_{k}^{\ast} \biggr]\biggr\},
\end{equation}
leading, after angular integration, to Eq. (\ref{00enmom}). The same type of calculation can be performed in the case of the 
expectation value of Eq. (\ref{ij1}). The result in this case can be written as 
\begin{eqnarray}
&& \langle T_{i}^{j}(\eta) \rangle = \frac{1}{a^4} \int \frac{d^{3}k}{(2\pi)^3} \biggl\{ P_{i}^{j}(k) 
\biggl[ |\Pi_{k}|^2 + \biggl(\frac{g'}{g}\biggr)^2
|{\cal A}_{k}|^2 + \frac{g'}{g}\biggl( {\cal A}_{k} \Pi_{k}^{\star} + {\cal A}_{k}^{\star}\Pi_{k} \biggr)\biggr]
\nonumber\\
&& - 4 k_{i} k^{j} |{\cal A}_{k}|^2 + \delta_{i}^{j}\biggl[ \biggl( k^2 - \biggl( \frac{g'}{g} \biggr)^2 \biggr) |{\cal A}_{k}|^2 - |\Pi_{k}|^2 
- \frac{g'}{g} \biggl(\Pi_{k}^{\ast} {\cal A}_{k} + \Pi_{k}{\cal A}_{k}^{\ast}\biggr) \biggr]\biggr\}.
\label{ij2}
\end{eqnarray}
If $f(k,\eta)$ is a generic scalar function of the modulus of the momentum and of the time the following 
identities hold:  
\begin{eqnarray}
&& \int d^{3}k k_{i} k^{j} f(k, \eta) = \frac{4 \pi}{3} \delta_{i}^{j} \int k^{5} f(k,\eta) d\ln{k},
\label{id1}\\
&& \int d^{3}k k^{i} f(k,\eta) =0.
\label{id2}
\end{eqnarray}
Using Eq. (\ref{id1})  into Eq. (\ref{ij2}) and recalling the explicit expression of the projector $P_{i}^{j}(k)$,  we can 
easily obtain
\begin{equation}
\langle T_{i}^{j} \rangle = - \frac{1}{3}\rho_{\rm r} \delta_{i}^{j} \equiv - p_{\rm r} \delta_{i}^{j},
\end{equation} 
as  can be argued from the fact that the classical energy--momentum tensor is traceless.
Finally, the expectation value $\langle T_{i}^{0} \rangle$ always vanishes by virtue of Eq. (\ref{id2}).
\end{appendix}

\end{document}